\documentclass[twocolumn,showpacs,preprintnumbers,amsmath,amssymb]{revtex4-2}
\usepackage{subcaption}
\usepackage{graphicx}
\usepackage{dcolumn}
\usepackage{bm}
\usepackage{hyperref}
\usepackage{amsfonts}
\usepackage{graphicx}

\begin{document}
\title{\textbf{Perturbation responses on topological synchrony in simplicial Kuramoto model}}

\author{Abhijeet Kumar}

\affiliation{
 Physics and Applied Mathematics Unit, Indian Statistical Institute, Kolkata 700108, India
}

\email{abhijeet.isic@gmail.com}
\email{palpalash93@gmail.com}
\email{dibakar@isical.ac.in}
\author{Palash Kumar Pal }
\affiliation{
 Physics and Applied Mathematics Unit, Indian Statistical Institute, Kolkata 700108, India
}

\author{Dibakar Ghosh}
 
\affiliation{
 Physics and Applied Mathematics Unit, Indian Statistical Institute, Kolkata 700108, India
}

\date{\today}

\begin{abstract}

Synchronization is conventionally understood as the emergence of a phase-locked collective state among interacting dynamical units. However, extending this notion to systems whose dynamical variables are defined on higher-dimensional simplices introduces fundamentally new constraints arising from the topology of the underlying simplicial complex. In the simplicial Kuramoto model, nontrivial topological cycles give rise to a higher dimensional harmonic subspace that is unaffected by the coupling and can therefore drift indefinitely, preventing a globally synchronized state. To resolve this we employ Hodge decomposition on simplicial Kuramoto dynamics and investigated the components.  Despite this topological obstruction, we show that the simplicial Kuramoto model admits fixed-point states in the exact and coexact sectors of the Hodge decomposition above critical coupling strengths, which we derive analytically for both sectors. This decomposition provides a generalized notion of synchronization in which the non-harmonic components converge to fixed states.  
We further investigate the robustness of these fixed-point states to external perturbations.
Excluding the drifting and non-interacting harmonic component, the perturbation response is governed by the spectrum of weighted Laplacian and recovers a Kirchhoff index dependence analogous to standard Kuramoto case. 
In contrast, the harmonic sector is non-dissipative, and consequently the fragility of the system increases with the dimension of this topological subspace. We characterize this effect numerically using triangulated tori with varying first Betti number and find a superlinear scaling relation between system fragility and the dimension of the harmonic sector.

\end{abstract}

\maketitle

\section{Introduction}
The canonical Kuramoto model \cite{10.1007/BFb0013365} remains one of the most widely studied  minimal models of synchronization \cite{acebron2005kuramoto},
  a phenomenon observed throughout nature,  from neuronal populations \cite{10.3389/fnhum.2010.00190,638513}, twinkling of fireflies \cite{638513}, audience applause \cite{PhysRevE.61.6987} and in engineered systems alike \cite{PhysRevLett.109.064101,PhysRevLett.89.054101}. 
A key limitation of dyadic network models, such as  the Kuramoto model, is that they fail to capture higher-order or group-level interactions among agents in the network topology. These higher-order interactions are increasingly recognized as essential for accurately modeling the structure and dynamics of real-world complex systems, from spreading processes \cite{Iacopini2019} to collective dynamics  \cite{PhysRevLett.121.228301, BOCCALETTI20231}. Group interactions are generically represented by two main mathematical models: hypergraphs and simplicial complexes \cite{battiston2020networks,bianconi2021higher}. 
Hypergraphs are more general, but the downward closure property of simplicial complexes endows them with an algebraic structure constrained by topology, enabling tools such as Hodge decomposition, which enormously helps us to dissect and understand the underlying dynamics of higher-order interactions. Hodge decomposition separates gradient from non-gradient flows, with applications ranging from statistical ranking \cite{Jiang2011} to the study of topological dynamical signals \cite{9044758}.  Topological signals \cite{barbarossa2020topological,torres2020simplicial},
are dynamical variables defined on simplexes of different dimensions for e.g., node signals are associated with $0$-simplexes, link signals with $1$-simplexes, and higher-dimensional signals with triangles or higher-order simplexes. They are also relevant in domains such as biological transport networks \cite{doi:10.1126/science.1177894,PhysRevE.103.062301}. Thus, for systems with higher-order interactions, simplicial complexes offer a principled framework to study collective dynamics such as diffusion \cite{Schaub:816290,PhysRevE.101.022308},  consensus \cite{PhysRevE.101.032310}, random walks \cite{doi:10.1137/18M1201019}, spreading \cite{Iacopini2019}, percolation \cite{PhysRevE.100.062311}, and evolutionary game theory \cite{Alvarez-Rodriguez2021}.
\\ 
The simplicial Kuramoto model \cite{PhysRevLett.124.218301} provides a natural extension of this framework to study collective behavior in higher-order networks. Unlike the node-based Kuramoto model, synchronization of dynamics defined on $k \geq 1$ simplices is fundamentally constrained by topology: nontrivial harmonic modes, associated with independent topological cycles, obstruct synchronization of the topological signal even at arbitrarily large coupling strength. A unified, topological perspective on synchronization across simplicial dimensions has so far remained largely undeveloped.
\\
While synchronization phenomena on simplicial complexes have received significant attention, the response of these higher-order dynamical systems to perturbations remains underexplored. In realistic systems, interactions are rarely isolated from external fluctuations, structural disorder, or localized disturbances. Therefore, understanding how perturbations propagate through higher-order network structures and how topology itself mediates the flow, localization, and persistence of disturbance becomes essential for characterizing the robustness, stability, and resilience of complex systems.   Perturbative analysis thus offers a window into the interplay between geometry, topology, and nonlinear dynamics in complex systems. \\
 In this article, we study the Hodge-decomposed simplicial Kuramoto model and prove that its exact and coexact components admit stable, fixed-point above an explicitly derived critical coupling strength. We show that this critical coupling strength depends only on the boundary operators and the natural-frequency content of the complex, and that the framework reduces exactly to the standard Kuramoto model at $k=0$.  Together, these results give a unified, topological characterization of synchronization across simplicial dimensions. \\
Next, we introduce perturbation in edge based Kuramoto model taking fixed phase state of coexact and exact components as our reference state and  analyze the response again via Hodge decomposition.  We take two types of perturbation,  box perturbation and sinusoidal perturbation applied for a finite time on edge based Kuramoto model. We derive the perturbation response analytically and compare with numerical results. We use simplicial configuration model \cite{PhysRevE.93.062311} to generate network topology to study the dynamics defined on edges.  The robustness of simplicial complex neglecting harmonic component under perturbation show dependence on Kirchhoff indices similar to standard Kuramoto model \cite{PhysRevLett.120.084101}. We then investigate how the perturbation response depends on topology by varying first Betti number ($\beta_1$) for dynamics defined on 1-cochain of a triangulated torus. This allows us to isolate the topological contribution, and we observe a scaling relation between the fragility  and $\beta_1$.

\section{Oriented simplicial complexes and Hodge theory}\label{sec2}
In this section, we briefly review the algebraic framework of  simplicial
complexes that are relevant to the subsequent study \cite{grady2010discrete,Hirani2003DiscreteEC,9044758}.
Sec.~\ref{sec21} defines oriented simplicial complexes and  Sec.~\ref{sec23} introduces cochains as the
topological signals  on which the dynamics take place. Sec.~\ref{sec24} defines
the boundary and coboundary operators that couple these signals across dimensions. These
operators combine in Sec.~\ref{sec25} to give the discrete Hodge Laplacian, whose kernel encodes the
topology of simplicial complex via the Betti numbers. Finally, Sec.~\ref{sec26} reviews the Hodge decomposition theorem, showing how topological signals can be uniquely decomposed into exact, harmonic, and coexact components that form the basis of the theoretical developments presented in the remainder of the paper.

\subsection{Oriented simplicial complexes}\label{sec21}
A $k$-simplex is an ordered set of $(k+1)$ vertices and is denoted by
$\sigma^k=[v_0,v_1,\dots,v_k]$. The orientation of a simplex is determined by the ordering of its vertices and are equivalent for even permutations.

A $k$-dimensional oriented simplicial complex  $\mathcal{K}$ is a set of simplices closed by inclusion where each simplex up to dimension $k$ is assigned an orientation \cite{grady2010discrete,Hirani2003DiscreteEC}. 
We write $N_k$ for the
number of $k-$simplices and define the set of oriented $k-$simplices as $\mathcal{S}_k=\{\sigma_i^k\}_{i=1}^{N_k}$. In the three-dimensional complexes
used here, $\sigma^0$ is a node, $\sigma^1$ a directed edge, $\sigma^2$ a triangle and $\sigma^3$ a tetrahedron where orientation is determined by ordering of its vertices.

\subsection{Cochains and topological signals}\label{sec23}
The real $k$-chain space $C_k(\mathcal{K};\mathbb{R})=\mathrm{span}_\mathbb{R}(\mathcal{S}_k)$ is the
vector space of formal real linear combinations of oriented $k$-simplices.
A $k$-cochain is a real-valued function acting on the space of $k$-chains. A $k$-cochain $\Phi_k\in C^k \cong \mathbb{R}^{N_k}$ assigns a real value to each oriented $k$-simplex,
$\Phi_k=(\Phi_k(\sigma_1^k),\dots,\Phi_k(\sigma_{N_k}^k))^\mathsf{T}$. \\
These cochains represent topological signals defined on simplices of a given dimension. In the context of Kuramoto dynamics, the cochain values correspond to variables associated with simplices.

\subsection{Boundary and co-boundary operators}\label{sec24}
The boundary operator $\partial_k: C_k\to C_{k-1}$ maps oriented $k$-simplex to oriented sums of their $(k-1)$-dimensional faces
\begin{equation}
\partial_k[v_0,\dots,v_k]=\sum_{i=0}^k (-1)^i [v_0,\dots,v_{i-1},v_{i+1},\dots,v_k].
\end{equation}
The matrix form of boundary operator is represented by the incidence matrix $
B_k
\in
\mathbb{R}^{N_{k-1}\times N_k},
$whose entries are given by
\begin{equation}
[B_k]_{ij}
=
\begin{cases}
+1,
&
\text{if }
\sigma_i^{k-1}
\text{ is coherently oriented with }
\sigma_j^k,
\\
-1,
&
\text{if }
\sigma_i^{k-1}
\text{ is incoherently oriented with }
\sigma_j^k,
\\
0,
&
\text{otherwise}.
\end{cases}
\end{equation}

The boundary operators satisfy the fundamental identity
$\partial_{k}\partial_{k+1} = 0$,which in matrix form can be written as
\begin{equation}
    B_{k}B_{k+1} = 0. 
    \label{bb=0}
\end{equation}

The $k$-th coboundary operator $\delta_k:C^k\to C^{k+1}$ is the transpose map, $\delta_k=B_{k+1}^\mathsf{T}$ induced by the $(k+1)$-th boundary operator.
The standard inner product between two $k$-cochains $\Phi_k,\Psi_k\in C^k$ can be written as
\begin{equation}
\langle \Phi_k,\Psi_k\rangle
=
\sum_{i=1}^{N_k}
\Phi_k(\sigma_i^k)\Psi_k(\sigma_i^k).
\end{equation}

\subsection{Hodge Laplacian and Betti Numbers}\label{sec25}
With these boundary and co-boundary operators, we can define discrete Hodge Laplacian \cite{doi:10.1137/18M1223101} that generalizes graph Laplacian to higher-order cochains as:
\begin{equation}
    L_k =  L^{down}_k + L^{up}_k = B^\mathsf{T}_k B_k + B_{k+1} B^\mathsf{T}_{k+1}.
    \label{Hl}
\end{equation}
Given the number of $k$ simplex in the complex is $N_k$, we can compute the corresponding Betti number  as $\beta_k = \operatorname{dim}(\operatorname{ker}(L_k)) = \operatorname{dim}(\operatorname{ker}(B_k)) - \operatorname{rank}(B_{k+1}) = N_k - \operatorname{rank}(B_k)-\operatorname{rank}(B_{k+1})$ \cite{ghrist2014elementary}.

\subsection{Hodge decomposition} \label{sec26}

A fundamental result of combinatorial Hodge theory states that the co-chain space admits the orthogonal decomposition \cite{Eckmann1944,Jiang2011}

\begin{equation}
    C^k \cong \mathbb{R}^{N_k} =  \operatorname{im} (B_{k}^\mathsf{T}) \oplus (\operatorname{ker}(B_k) \cap  \operatorname{ker}(B_{k+1}^\mathsf{T})) \oplus \operatorname{im} (B_{k+1}).
    \label{hodge decomposition}
\end{equation}

Consequently, any $k-$ cochain $x_k \in C^k$ can be decomposed as
\begin{equation}
x_k
=
x_{k,E}
+
x_{k,H}
+
x_{k,C}
= B_k^\mathsf{T}z_{k-1}  + x_{k,H} + B_{k+1}z_{k+1},
\label{kcochaindecomposition}
\end{equation}
where,

\begin{itemize}
    \item $x_{k,E}\in\operatorname{im}(B_k^\mathsf{T}) \implies x_{k,E} = B_k^\mathsf{T}z_{k-1} $ for some $(k-1)$ co-chain and referred to as exact component,

    \item $x_{k,H}\in (\operatorname{ker}(B_k) \cap  \operatorname{ker}(B_{k+1}^\mathsf{T})) $ is the harmonic component, and 

    \item $x_{k,C}\in\operatorname{im}(B_{k+1}) \implies x_{k,C} =  B_{k+1}z_{k+1} $ for some $(k+1)$ co-chain and referred to as coexact component.
\end{itemize}

The exact component corresponds to gradient-like topological flows, whereas the coexact component corresponds to curl-like circulations induced by higher-order interactions. The harmonic component represents topological modes that are left out and thus is simultaneously divergence-free and curl-free.

The subspace decomposition guarantees that components satisfy the orthogonality relations:
\begin{equation}
    B_k x_{k,C}=0,
\qquad
B_{k+1}^\mathsf{T} x_{k,E}=0,
\label{orth1}
\end{equation}

together with
\begin{equation}
B_k x_{k,H}=0,
\qquad
B_{k+1}^\mathsf{T} x_{k,H}=0.
\label{orth2}
\end{equation}

Thus, harmonic modes are invisible to both lower and upper boundary projections and are entirely determined by the topology of the simplicial complex.

Starting from a $k$-cochain, we can get the Hodge components by exploiting the orthogonality relations together with the fact that boundary of boundary is null \eqref{bb=0}.  By multiplying $B_k$ and $B_{k+1}^\mathsf{T}$ to \eqref{kcochaindecomposition} sequentially, we get exact component as  $ x_{k,E} = B_k^\mathsf{T}  (B_kB_k^\mathsf{T})^+ B_kx_k$ and coexact component as  $x_{k,C} = B_{k+1}(B_{k+1}^\mathsf{T}B_{k+1})^+B_{k+1}^\mathsf{T} x_k$. Here, $(B_kB_k^\mathsf{T})^+$ denotes pseudoinverse \cite{ben2006generalized} of $B_kB_k^\mathsf{T}$.

\section{Simplicial Kuramoto dynamics}\label{sec3}
In this section, we consider the simplicial Kuramoto model and analyze its synchronization properties through the Hodge decomposition framework. Sec.~\ref{sec31} decomposes the dynamics into exact, harmonic, and coexact sectors, thereby decoupling the evolution into independent topological subspaces. Sec.~\ref{sec32} derives the critical coupling strengths for the existence of  fixed-point states of exact and coexact components. Sec.~\ref{sec33} shows how the proposed framework is coherent with the notion of synchronization for classical Kuramoto model. Sec.~\ref{sec34} extends the analysis to the edge-based higher-order Kuramoto model and generalizes the notion of synchronization from a topological perspective. Finally, Sec.~\ref{sec35} investigates the linear stability of the fixed points in the exact and coexact sectors.
\par We consider the higher-order Kuramoto model \cite{PhysRevLett.124.218301} defined on each simplex $\alpha$ of dimension $k$ of the simplicial complex. Let $\theta_{(k)} \in C^k$ 
be a $k$-cochain representing the phases associated with oriented $k$ simplex.

The simplex-based higher-order Kuramoto model is given by
\begin{equation}
\label{eq:higher_order_kuramoto2}
\dot{\theta}_{(k)}
=
\omega
-
\sigma_d B_k^\mathsf{T} \sin(B_k\theta)
-
\sigma_u B_{k+1} \sin(B_{k+1}^\mathsf{T}\theta),
\end{equation}
where,

\begin{itemize}
    \item $\omega \in C^k$ is the vector of natural frequencies associated with each simplex of dimension $k$,

    \item $B_k^\mathsf{T}\sin(B_k\theta)$ represents lower-adjacent interactions induced through $(k-1)$ simplexes with coupling strength $\sigma_d$,

    \item $B_{k+1}\sin(B_{k+1}^\mathsf{T}\theta)$ represents upper-adjacent interactions induced through superface of associated simplex with coupling strength $\sigma_u$.
\end{itemize}
From \eqref{eq:higher_order_kuramoto2}, we recover the classic Kuramoto model by observing the dynamics on $k=0$ in which the interaction is taking place via $k=1$ simplex, i.e., via edges and   is given by 
\begin{equation}
\label{eq:classical_kuramoto2}
\dot{\theta}
=
\omega
-
\sigma B_1 \sin(B_1^\mathsf{T} \theta).
\end{equation}
Despite the oriented incidence matrix $B_1$, the interaction term is orientation invariant and thus for $k=0$, simplicial Kuramoto model coincide with the undirected Kuramoto model.

\subsection{Hodge decomposed simplicial Kuramoto dynamics}\label{sec31}
Hodge decomposition of cochains can be used to separate the simplicial Kuramoto dynamics into exact, harmonic, and coexact sectors \cite{arnaudon2022connecting}.
By Hodge decomposition of the topological signal,
\begin{equation}
\theta
=
\theta_E
+
\theta_H
+
\theta_C,
\end{equation}
which separates phases into exact, coexact, and  harmonic components.

Similarly, the natural frequency vector admits the decomposition
\begin{equation}
\omega
=
\omega_E
+
\omega_H
+
\omega_C.
\end{equation}

Using the orthogonality relations \eqref{orth1} and \eqref{orth2}, 
the nonlinear dynamical equation \eqref{eq:higher_order_kuramoto2} decouple into independent dynamics \cite{nurisso2024unified} in three Hodge subspaces: 

\begin{subequations}
\begin{align}
\dot{\theta}_E &=
\omega_E- \sigma_d B_k^\mathsf{T} \sin(B_k\theta_E), \label{hodgeexact} \\
\dot{\theta}_H &= \omega_H, \label{hodgeharmonic} \\ 
\dot{\theta}_C &= \omega_C - \sigma_u B_{k+1} \sin(B_{k+1}^\mathsf{T}\theta_C). \label{hodgecoexact}
\end{align}
\label{hodgedecomposed}
\end{subequations}

Eqs.~\eqref{hodgeexact} and~\eqref{hodgecoexact} describe the dissipative exact and coexact sectors, whereas Eq.~\eqref{hodgeharmonic} governs the harmonic topological modes.

\subsection{Coexact and exact components}\label{sec32}
We can substitute $\theta_C = B_{k+1} \psi_C$ in Eq.~\eqref{hodgecoexact}, where $\psi_{C}$ is some $(k+1)$ cochain, 
\begin{equation}
    \begin{aligned}
         B_{k+1} \dot\psi_C = \omega_C - \sigma_u B_{k+1} \sin(B_{k+1}^\mathsf{T} B_{k+1} \psi_C).    
    \end{aligned}
    \label{C_red_dyn}
\end{equation}
Multiplying both sides by $B_{k+1}^+ = (B_{k+1}^\mathsf{T} B_{k+1})^{+}B_{k+1}^\mathsf{T}$ where, $B_{k+1}^+$ is Moore- Penrose pseudoinverse, 
\begin{equation}
    B_{k+1}^+ B_{k+1}\dot\psi_C = B_{k+1}^+ \omega_C - \sigma_u B_{k+1}^+ B_{k+1} \sin(B_{k+1}^\mathsf{T}B_{k+1}\psi_C)
    \label{eq5}
\end{equation} 

Since $C^{k+1} = \operatorname{im}(B^\mathsf{T}_{k+1}) \oplus \operatorname{ker}(B_{k+1})$, so the choice of $\psi_C $ is not unique and we can choose minimum norm representative which satisfies $\psi_C \in \operatorname{im}(B^\mathsf{T}_{k+1})$.
Denoting $B_{k+1}^+ B_{k+1} = P_{\operatorname{im}(B^\mathsf{T}_{k+1})}$, where $P_{\operatorname{im}(B^\mathsf{T}_{k+1})}$ is orthogonal projector onto $\operatorname{im}(B_{k+1}^\mathsf{T})$. Therefore, $P_{\operatorname{im}(B^\mathsf{T}_{k+1})}\psi_C = \psi_C$. This gives,

\begin{equation}
   \dot\psi_C = B_{k+1}^+ \omega_C - \sigma_u P_{\operatorname{im}(B^\mathsf{T}_{k+1})}\sin(B_{k+1}^\mathsf{T}B_{k+1}\psi_C).
    \label{5.1}
\end{equation}

 We can define a positive definite matrix $M = B_{k+1}^\mathsf{T} B_{k+1}$ and $\nu = M^{+}B_{k+1}^\mathsf{T} \omega_C$. Then our dynamics reduces to
\begin{equation}
    \dot\psi_C = \nu - \sigma_u P_{\operatorname{im}(B^\mathsf{T}_{k+1})} \sin (M \psi_C).
    \label{eq6}
\end{equation}

Since  $\nu $ and $ M\psi_C $ lies in same subspace, i.e, in $\operatorname{im}(B_{k+1}^\mathsf{T})$, both the natural frequency and coupling term act within the same subspace allowing them to cancel each other at a fixed point.   Also, $|P_{\operatorname{im}(B^\mathsf{T}_{k+1})} \sin (M \psi_C)| \leq  |  \sin (M \psi_C)|$, component-wise.  Thus, the necessary condition can  be written as   $ |\dfrac{\nu_k}{\sigma_u}| \leq 1, \hspace{0.2cm} \forall k $. This gives us a critical coupling strength for which $ \psi_C $ have a solution. Thus  the coexact component $ B_{k+1}\psi_C $ goes into fixed phase state (since $B_{k+1}$ is time-independent transformation) as
\begin{equation}
\sigma_u \geq \sigma_{crit,u} = ||\nu||_{\infty} = || (B_{k+1}^\mathsf{T}B_{k+1})^{+}B_{k+1}^\mathsf{T} \omega ||_\infty , 
\label{coexact critical}
\end{equation}
where $|x||_\infty$ represents the maximum component value of vector $x$.


Similarly, for the exact case, $ \theta_E = B^\mathsf{T}_k\Phi_E$ where, $\Phi_E$ is some $(k-1)$ cochain; dynamics can be written as 
\begin{equation}
    B_k^\mathsf{T}\Phi_E = \omega_E - \sigma_d B_k^\mathsf{T}\sin(B_kB_k^\mathsf{T}\Phi_E)
\label{eq8}
\end{equation}

Multiplying both sides by the Moore - Penrose pseduoinverse, $B_k^{\mathsf{T}+} = (B_kB_k^\mathsf{T})^{+}B_k$, we get
\begin{equation}
    B_k^{\mathsf{T}+} B_k^\mathsf{T}\dot\Phi_E = B^{\mathsf{T}+}_k\omega_E - \sigma_d B_k^{\mathsf{T}+} B_k^\mathsf{T}\sin(B_kB_k^\mathsf{T}\Phi_E)
\end{equation}
Since we can decompose $(k-1)$ cochain space as $\mathbb{R}^{N_{k-1}} = \operatorname{im}(B_k) \oplus \operatorname{ker}(B^\mathsf{T}_k)$ and denoting $B_k^{\mathsf{T}+}B^\mathsf{T}_k = P_{\operatorname{im}(B_k)}$, where $P_{\operatorname{im}(B_k)}$ is orthogonal projection operator onto subspace $\operatorname{im}(B_k)$. Considering minimum-norm representative that satisfies $\Phi_E \in \operatorname{im}(B_k)$, we get $P_{\operatorname{im}(B_k)} \Phi_E = \Phi_E$. So, our equation becomes
\begin{equation}
    \dot\Phi_E = B^{\mathsf{T}+}_k\omega_E - \sigma_d P_{\operatorname{im}(B_k)} \sin(N\Phi_E).
\end{equation}

Defining $N = B_kB_k^\mathsf{T}$ and $ \mu =  (B_kB_k^\mathsf{T})^{+}B_k\omega_E$,  the dynamics in exact subspace reduces to 
\begin{equation}
    \dot\Phi_E = \mu - \sigma_d P_{\operatorname{im}(B_k)} \sin(N\Phi_E).
    \label{eq9}
\end{equation}
By analogous argument as in coexact case, for $|\dfrac{\mu_k}{\sigma_d}| \leq 1, \hspace{0.1cm} \forall k$,  we will get a fixed phase state criterion for exact phase
\begin{equation}
\sigma_d \geq \sigma_{crit,d} = ||\mu||_{\infty} = ||  (B_kB_k^\mathsf{T})^{+}B_k \omega ||_\infty .
\label{exact critical}    
\end{equation}
\\

\subsection{Classic Kuramoto model} \label{sec33}
The Hodge separation of dynamics allows rewriting of Eq.~\eqref{eq:classical_kuramoto2} as 
\begin{subequations}
\begin{align}
\dot{\theta}_H &= \omega_H, \label{eq:harmonic_classical} \\
\dot{\theta}_C &= \omega_C - \sigma B_1 \sin(B_1^\mathsf{T} \theta_C). \label{eq:coexact_classical}
\end{align}
\label{classical}
\end{subequations}

Eq.~\eqref{eq:harmonic_classical} describes rigid collective rotation of the harmonic sector with constant collective frequency. In contrast, the Eq.~\eqref{eq:coexact_classical} governs the diffusive synchronization dynamics.

\begin{figure*}[!ht]
    \centering
    \includegraphics[width=0.8\linewidth]{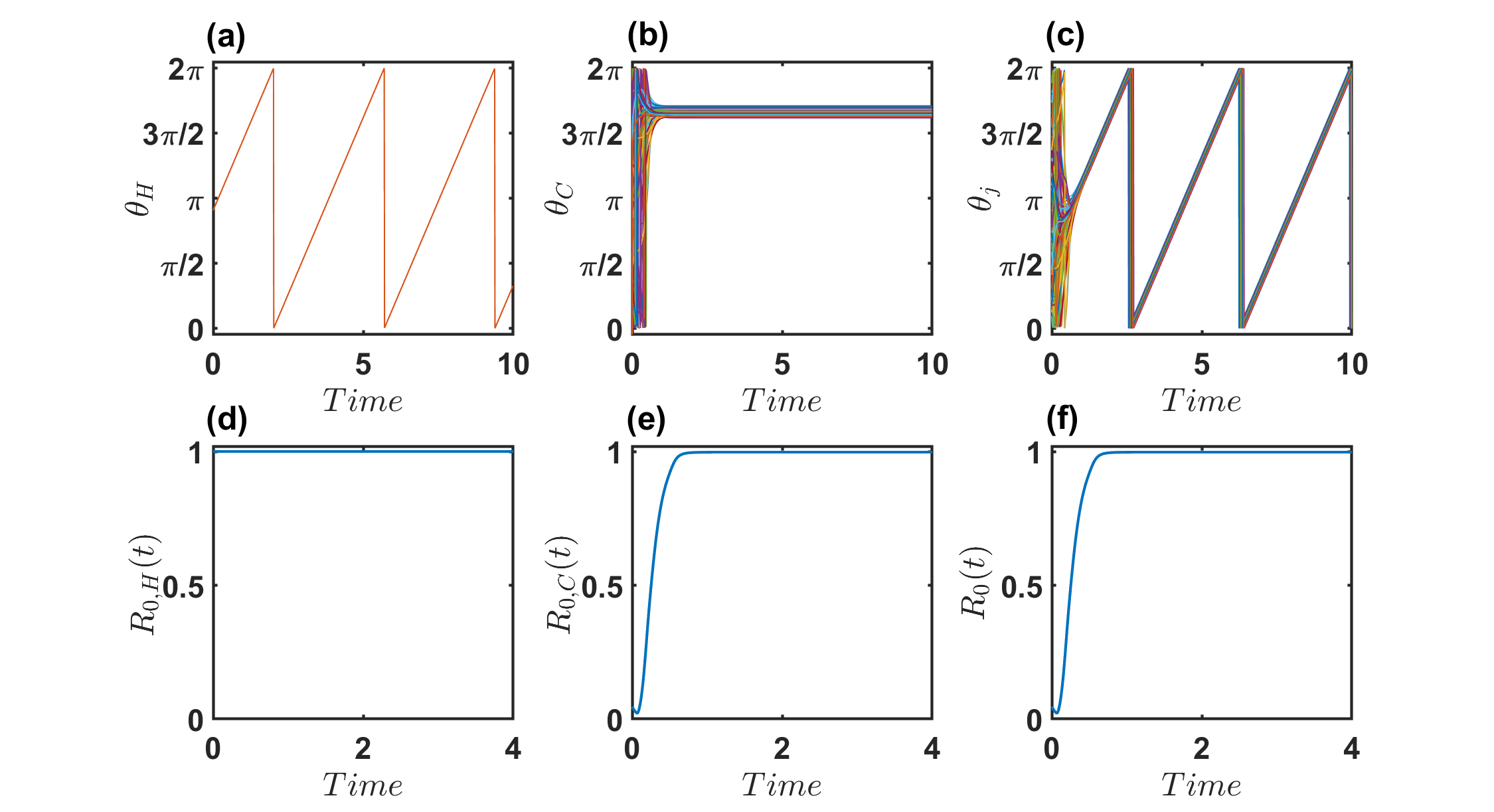}
    \caption{\textbf{Synchronization transition of node phases, and its two components in the classic Kuramoto model.} The upper row represents the time series of (a) the harmonic component vector $\theta_H$, (b) the coexact component vector $\theta_C$, and (c) the nodes vector or node signal $\theta$. Panels (d), (e), and (f) refer, respectively, to their order parameters corresponding to the vectors $\theta_H$, $\theta_C$, and $\theta$. The coupling strength is $\sigma=4.0$, taken above the critical value. The network is the 1-skeleton of a two-dimensional configuration model of simplicial complex with $100$ nodes, $m=2$, and $\gamma=2.6$. }
    \label{fig:timeseries_std}
\end{figure*}

\begin{figure}[htbp]
\centering

\includegraphics[width=0.7\linewidth]{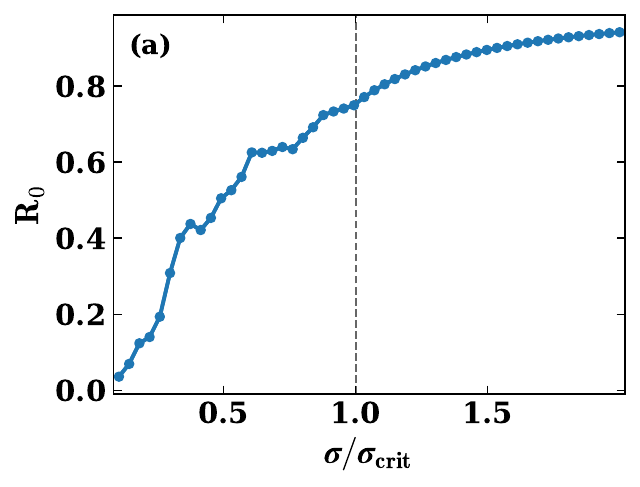}

\vspace{0.5em}
\includegraphics[width=0.7\linewidth]{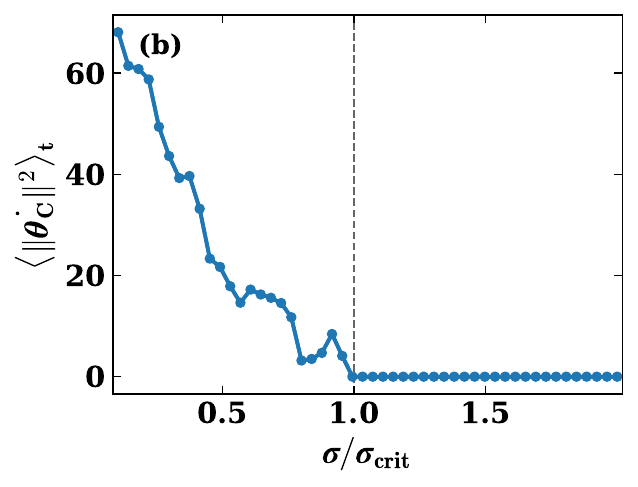}

\caption{Standard Kuramoto model defined on 1-skeleton of a two-dimensional configuration model of simplicial complex with $200$ nodes, $m=2$, and $\gamma=2.6$ and $\omega \sim \mathbf{U}(1,3)$. (a) Order parameter, $R$ is plotted against the the normalized coupling strength. (b) The  averaged $\dot \theta_C$ after transients showing onset of synchronization as coexact sector freezes. }
\label{fig:stdkurmotovscoupling}
\end{figure}

Here, the dynamics is defined on one dimensional simplicial complex, i.e., regular connected graph which has 1 dimensional Harmonic component as $\beta_1 = 1.$ Thus, $\omega_H$ will be mean of the  distribution of $\omega$. Beyond the derived critical coupling strength in Eq.~\eqref{coexact critical}, i.e., $\sigma > ||(B_{1}^\mathsf{T}B_{1})^{-1}B_{1}^\mathsf{T} \omega ||_\infty$, we get frozen coexact dynamics while the phase is rotating with a constant harmonic frequency. We plot the dynamics of standard Kuramoto model defined on 1-skeleton of simplicial complex in Fig.~\ref{fig:timeseries_std}. The harmonic component is one-dimensional, while the coexact component exhibits a phase-locked state above the critical coupling strength. Together, these reproduce the classical synchronization picture: phases drift collectively while remaining mutually phase-locked.
We  plot order parameter $R$ vs $\sigma/\sigma_{crit}$ and time-averaged velocity, $||\dot\theta_C||^2$ vs $\sigma/\sigma_{crit}$ in Fig.~\ref{fig:stdkurmotovscoupling} on 1- skeleton of the simplicial complex. As the coupling strength increases, the coexact component saturates to a fixed point, and correspondingly the order parameter approaches 1 as the phases synchronize.
\\

The critical coupling strength derived here,  $\sigma > ||(B_{1}^\mathsf{T}B_{1})^{+}B_{1}^\mathsf{T} \omega ||_\infty$  is dependent on network topology and valid for finite number of oscillators.

Therefore, at synchronization $\theta_{sync}(t) = \theta_H + \theta_C =  \langle \omega \rangle t + \theta^\ast $, where $\theta^\ast$ is offset caused by initial condition and frozen coexact dynamics.  This coincides with our classic notion of synchronization.

\subsection{Edge based higher order Kuramoto model}\label{sec34}
We now consider dynamics defined on edges, i.e., on  1-cochains. We can write the Hodge decomposed dynamics from \eqref{hodgedecomposed} as
\begin{subequations}
\begin{align}
\label{eq:exact_dynamics}
\dot{\theta}_E &= \omega_E -
\sigma_d B_1^\mathsf{T} \sin(B_1\theta_E), \\ 
\label{eq:harmonic_dynamics}
\dot{\theta}_H
&=
\omega_H, \\
\label{eq:coexact_dynamics}
\dot{\theta}_C &=
\omega_C -
\sigma_u B_2 \sin(B_2^\mathsf{T}\theta_C).
\end{align}
\label{edgebased}
\end{subequations}

\begin{figure}[htbp]
\centering

    \centering
    \includegraphics[width=0.7\linewidth]{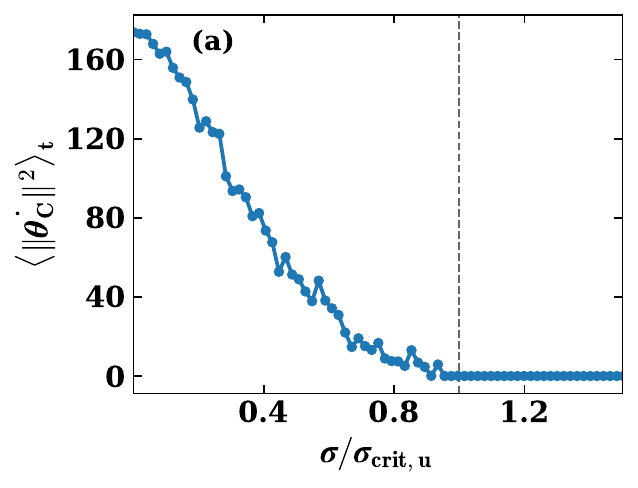}

    \vspace{0.5em}\includegraphics[width=0.7\linewidth]{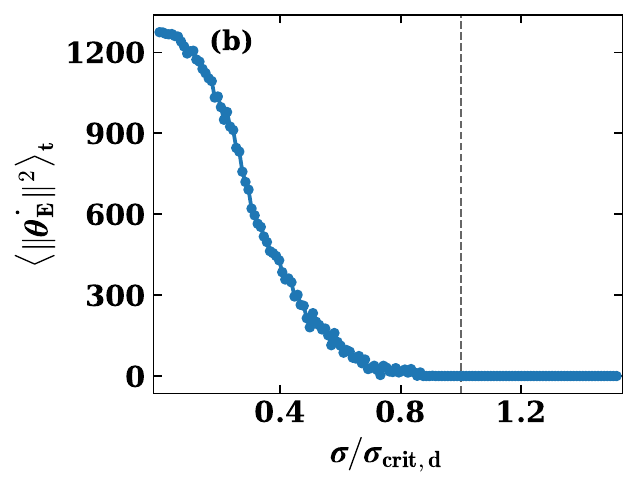}

\caption{Simplicial  Kuramoto model on configuration  with power-law exponent $\gamma=2.5$ and $m=2$ having $N_0=200, N_1 =856$ and $N_2=344$. $\omega \sim \mathbf{U}(1,3)$.  Time averaged coexact component velocity plotted on normalized coexact critical coupling strength (a) and time averaged exact component velocity on normalized exact critical coupling strength(b). }
\label{fig:kurmotovscoupling_edge}
\end{figure}

The harmonic component though nontrivial is decoupled from the interaction and thereby evolves solely on the natural frequency. The coexact and exact sectors show frozen phase state beyond their respective critical coupling strength.  
We plot time-averaged velocity for coexact $||\dot\theta_C||^2$ and exact $||\dot\theta_E||^2$ vs $\sigma/\sigma_{crit}$ in 
Fig~\ref{fig:kurmotovscoupling_edge} for configuration simplicial network. Thus in classical sense the phases themselves do not synchronize but the Hodge decomposed phases behave similarly to what we have seen in the case of classical Kuramoto.

\begin{figure*}[!ht]
    \centering
    \includegraphics[width=0.8\linewidth]{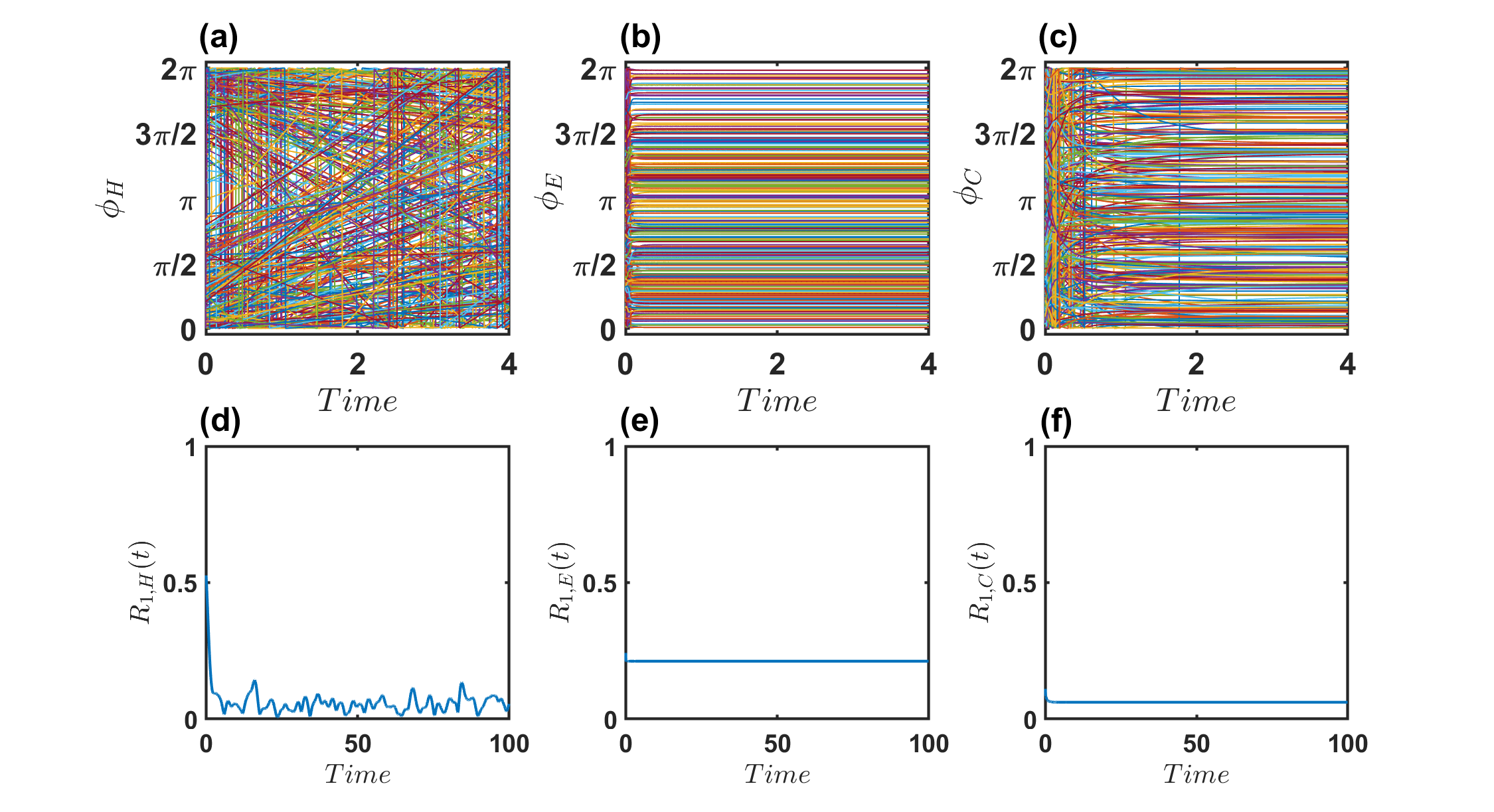}
    \caption{\textbf{Synchronization transition of the Hodge sectors of the edge signal in the higher-order Kuramoto model} The upper row represents the time series of (a) the harmonic component vector $\theta_H$, (b) the exact component vector $\theta_E$, and (c) the coexact component vector $\theta_C$. Panels (d), (e), and (f) refer, respectively, to their order parameters corresponding to the vectors $\theta_H$, $\theta_E$, and $\theta_C$. The coupling strengths are taken above the critical values. The higher-order network structure is a two-dimensional configuration model of simplicial complex of $100$ nodes with $m=2$, and $\gamma=2.6$.}
    \label{fig:timeseries_edge}
\end{figure*}

Thus in case of higher order dynamics the notion of synchronization is overgeneralized and a Hodge-centric viewpoint reveals a much unifying viewpoint. As in case of classic Kuramoto we have seen the non-harmonic sector freezes beyond a certain critical coupling strength, the analogous phenomena occurs in higher order Kuramoto model as coexact and exact sectors freezes. We plot the edge dynamics in Fig.~\ref{fig:timeseries_edge} which shows incoherently drifting harmonic components and frozen phase state of coexact and exact components beyond the critical coupling strength. Therefore, this viewpoint serves as more adequate notion of synchronization particularly from topological perspective. \\
The global phase shift symmetry in standard Kuramoto model, $\theta \rightarrow \theta + \alpha \mathbf{1}$ also generalizes naturally to simplicial setting. Since, the dynamics depends on $\theta$ through $B_{k+1}^\mathsf{T} \theta$ and $B_k \theta$ any phase shift $\theta \rightarrow \theta + h$ for $h \in \operatorname{ker}(L_k)$ leaves the equation of motion invariant. The symmetry group generalizes from $\mathbb{R}$, generated by $\mathbf{1} \in \operatorname{ker}(L_0)$ in case of standard Kuramoto model to $\mathbb{R}^{\beta_k}$ generated by basis of $\operatorname{ker}(L_k)$ with first Betti $\beta_k$ denoting the dimension of the symmetry group.
\\

This separation of dynamics into  dissipative and topological Hodge sectors forms the basis for the perturbation and fragility analysis developed in the subsequent sections.

\subsection{Stability of fixed point of Coexact and Exact components}\label{sec35}
For $\sigma > \sigma_{crit,d,u} $, where $\sigma = \sigma_d = \sigma_u$ coexact and exact sectors admits fixed points. Therefore, in case of coexact fixed phase state  $\theta_C(t) =  \theta_C^\ast$ and this state is a solution of Eq.~\eqref{hodgecoexact}, satisfying 
\begin{equation}
    \dot\theta_C^\ast = 0 =  \omega_C - \sigma B_{k+1} \sin\left(B_{k+1}^\mathsf{T}\theta_C^\ast\right). 
    \label{eq13}
\end{equation}

Taking a small deviation around the fixed point, $\theta_C^\ast \rightarrow \theta^\ast_C + \delta \theta_C$ Eq.~\eqref{hodgecoexact} becomes 

\begin{equation}
\begin{aligned}
     \delta\dot\theta_C = \omega_C -  \sigma B_{k+1}\sin(B_{k+1}^\mathsf{T}(\theta_C^\ast + \delta\theta_C))
\end{aligned}
\label{eq13a}
\end{equation}

We can approximate the sine term in the above equation, since $B^\mathsf{T}_{k+1}\delta\theta_C \ll 1$, as
\begin{equation}
\begin{aligned}
    \sin(B_{k+1}^\mathsf{T}(\theta_C^\ast + \delta\theta_C) &\approx \sin(B_{k+1}^\mathsf{T}\theta_C^\ast) + \\ & \operatorname{diag}(\cos(B_{k+1}^\mathsf{T} \theta_C^\ast))B_{k+1}^\mathsf{T}\delta \theta_C.
\end{aligned} 
    \label{eq13b}
\end{equation}
Under the constraint that $\cos(B^\mathsf{T}_{k+1} \theta_C^\ast)_i > 0$, i.e., $|B^\mathsf{T}_{k+1}\theta_C^\ast|_i < \pi/2$ component-wise,  we can define a weighted up Laplacian,  $\tilde L^{up}_k = \sigma B_{k+1}\operatorname{diag}(\cos(B_{k+1}^\mathsf{T} \theta_C^\ast))B_{k+1}^T$.
Since the weights are positive, $\tilde L^{up}_k$ will be positive semi-definite. Thus, our deviation equation reduces to 
\begin{equation}
    \delta\dot\theta_C = - \tilde L^{up}_k\delta \theta_C. 
    \label{eq13c}
\end{equation}
This is a Laplacian type flow and is exponentially stable \cite{Strogatz1991} in $\operatorname{im}(B_{k+1})$, i.e., in coexact sector. 
\\

Similarly, we can write exact dynamics under fixed phase state  $\theta_E(t) =  \theta_E^\ast$ and this state  satisfies,

\begin{equation}
 \dot{\theta}^\ast_E = 0=\omega_E - \sigma_d B_k^\mathsf{T} \sin(B_k\theta^\ast_E).
 \label{eq14}
\end{equation}
Taking small deviation around the fixed point, $\theta_E^\ast \xrightarrow{} \theta_E^\ast + \delta\theta_E $, Eq.~\eqref{hodgeexact} becomes
\begin{equation} 
     \delta\dot\theta_E = \omega_E -  \sigma B_{k}^\mathsf{T} \sin(B_{k}(\theta_E^\ast + \delta\theta_E))
\label{eq14a}
\end{equation}
In this case also, we can approximate sine term as $B_k \delta \theta_E \ll 1$, with 
\begin{equation}
    \sin(B_{k}(\theta_E^\ast + \delta\theta_E) \approx \sin(B_{k}\theta_E^\ast) + \operatorname{diag}(\cos(B_{k} \theta_E^\ast))B_{k}\delta \theta_E.
    \label{eq14b}
\end{equation}
For $\cos(B_{k} \theta_E^\ast)_i > 0$, i.e., $|B_{k}\theta_E^\ast|_i < \pi/2$ component-wise,  we can define a weighted down Laplacian,  $\tilde L^{down}_k = \sigma B_{k}^\mathsf{T}\operatorname{diag}(\cos(B_{k} \theta_E^\ast))B_{k} $.
 The weights here, are also positive and  $\tilde L^{down}_k$ will be positive semi-definite. Thus, our deviation equation reduces to 
\begin{equation}
    \delta\dot\theta_E = - \tilde L^{down}_k\delta \theta_E .
    \label{eq14c}
\end{equation}
This is also a Laplacian type flow and is exponentially stable in exact sector, i.e.,  in $\operatorname{im}(B^\mathsf{T}_{k})$.

 \subsection{Synchronization of projected dynamics}
  In recent works \cite{PhysRevLett.124.218301,Ghorbanchian2021}, it is studied that the projected phases synchronize, i.e.,  $ \psi_u= B^\mathsf{T}_{k+1} \theta $ and $\psi_d= B_k \theta $. From  orthogonality relations \eqref{orth1},\eqref{orth2}, we get $B^\mathsf{T}_{k+1} \theta \iff B^\mathsf{T}_{k+1}\theta_C$ and $B_k \theta \iff B_k \theta_E$.  This is consistent with the analysis suggested here. As $\theta_C$ and $\theta_E$ goes into fixed point ; its projection via incidence operators collapses them into a single value giving synchronization of projected variables. Therefore, referring synchronization in higher order Kuramoto model to synchronization of projected observable is equivalent to freezing of coexact and exact components beyond the critical coupling strength.


\section{Perturbation Analysis}\label{sec4}
\begin{figure*}[!ht]
    \centering
    \includegraphics[width=0.8\linewidth]{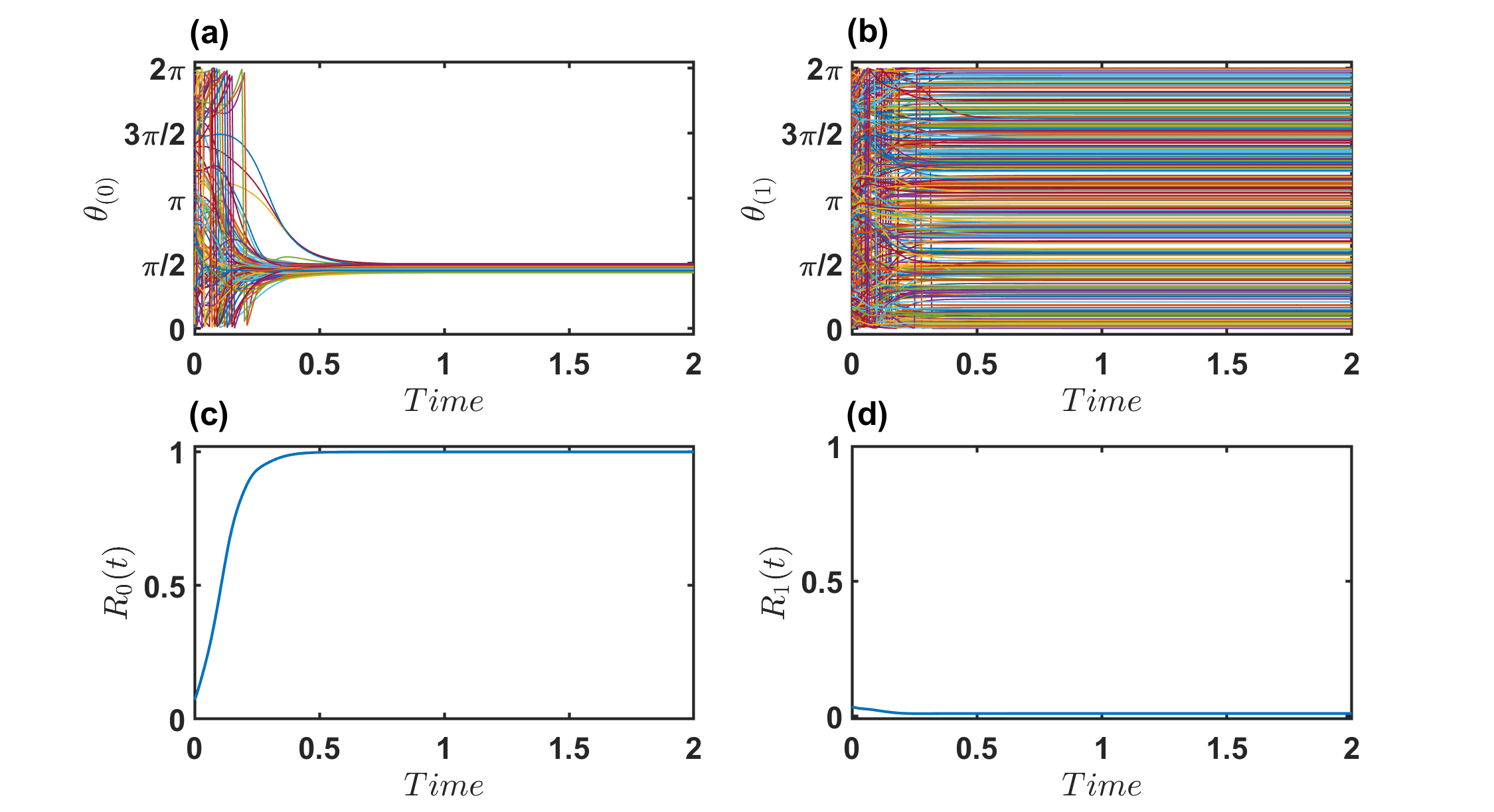}
    \caption{\textbf{Synchronization transitions of node and edge signals in a configuration model of a two-dimensional simplicial complex with $100$ nodes, $m=2$, and $\gamma=2.6$.} Panels (a) and (b) show the time evolution of the node and edge phases, respectively. The corresponding synchronization order parameters for the node and edge signals are presented in panels (c) and (d), respectively. The coupling strengths are fixed at $\sigma_u=\sigma_d=6.0$, which is above the critical synchronization threshold. The natural frequencies of both the node and edge signals are selected from subspaces in which their harmonic components vanish.}
    \label{fig:fixedpoint}
\end{figure*}

The Hodge subspaces' evolution is independent, and we  use the static phase state of non-harmonic sectors as reference state to analyze the perturbation and analogously call it the synchronized state from a topological perspective. In case of classical Kuramoto model (simplicial Kuramoto model defined for k=0), a common practice is to go into co-rotating frame \cite{Kuramoto1984} which is essentially removing Harmonic part of dynamics which is one dimensional for $k=0$, thereby, effectively reducing the dynamics to only coexact component. In higher order dynamics, the Harmonic part is non-trivial and the concept of co-rotating frame does not generalize. However, the analogous operation is removing harmonic component for any $k \geq 1$. In Fig.~\ref{fig:fixedpoint} we plot the dynamics of Kuramoto model defined on nodes, $\theta_{(0)}$  and $\theta_{(1)}$ with removed natural harmonic frequency, i.e., $\omega_H =0$. The dynamics settle into fixed phase state for $\sigma > \sigma_{crit,u,d}$ for node and edge dynamics. 





However, we treat every Hodge component as it gives the complete analytic picture. 
Introducing perturbation $\delta P (t)$ in Eq.~\eqref{eq:higher_order_kuramoto2}, where $\delta P (t)$ is $k$-cochain and we can decompose it as $\delta P(t) = \delta P_C(t) + \delta P_H(t) + \delta P_E(t)$ using Hodge decomposition \eqref{hodge decomposition}.  Thus, the perturbed decomposed dynamical equation have an extra driving  term in Eq.~\eqref{hodgedecomposed}.

\subsection{Coexact and Exact sectors}
Coexact component under introduction of perturbation  can be written as its deviation from static state as  $\theta_C^\ast(t) \rightarrow \theta_C^\ast + \delta\theta_C(t) $. From Eq.~\eqref{hodgecoexact}, we get evolution equation around perturbation as 
\begin{equation}
\begin{aligned}
     \delta\dot\theta_C = \omega_C -  \sigma B_{k+1}\sin(B_{k+1}^\mathsf{T}(\theta_C^\ast + \delta\theta_C(t))) + \delta P_C(t).
\end{aligned}
\label{eq111}
\end{equation}
From, Eqs.~\eqref{eq13b} and \eqref{eq13c}, we can write the above equation using $\tilde L^{up}_k$ as 

\begin{equation}
    \delta\dot\theta_C = \delta P_C- \tilde L^{up}_k\delta \theta_C.
    \label{eq141}
\end{equation}

Similarly, for exact case under the introduction of perturbation $\delta P_E$ in static state as $\theta_E^\ast \rightarrow \theta_E^\ast + \delta \theta_E(t)$ and our evolution Eq.~\eqref{hodgeexact} becomes:
\begin{equation}
    \delta\dot\theta_E = \omega_E -  \sigma B_{k}^\mathsf{T}\sin(B_{k}(\theta_E^\ast + \delta\theta_E(t))) + \delta P_E.
\label{eq11}
\end{equation}
Using, Eqs.~\eqref{eq14b} and \eqref{eq14c} we can write the above equation as 

\begin{equation}
     \delta\dot\theta_E = \delta P_E- \tilde L^{down}_k\delta \theta_E .
     \label{15}
\end{equation}

\subsection{Harmonic sector}
The harmonic component \eqref{hodgeharmonic} drifts without any interaction. To measure deviation of harmonic component, we redefine phase as $\xi_H(t) = \theta_H(t) - \theta_{H}^\ast - \omega_H t$, where $\theta_{H}^\ast$ is initial phase offset. This does not change the structure of dynamics since $ \xi_H \in \operatorname{ker}(B_k) \cap  \operatorname{ker}(B_{k+1}^\mathsf{T}))$. 
Now the deviation equation for harmonic component simply becomes 
\begin{equation}
    \dot\xi_H = \delta P_H(t).
    \label{16}
\end{equation}

The harmonic sector is inherently non-dissipative so the perturbation energy deposited into $\beta_k$, harmonic modes is retained as quasi-periodic motion on $\mathbb{T}^{\beta_k}$ for generic $\omega_H$ or $\delta P_H$. For projected dynamics,
$B^\mathsf{T}_{k+1}\theta$ and $B_k \theta$, its completely invisible and it represents a dynamically neutral degree of freedom whose dimension equals $\beta_k$.

\subsection{Putting it together}
The deviation can now be expressed as $ \delta \theta = \delta \theta_C + \delta \theta_E + \xi_H $. Therefore, the response of perturbation for uncompounded $\theta$ is given by 
\begin{equation}
    \delta \dot \theta = \delta P_C(t)- \tilde L^{up}_k\delta \theta_C + \delta P_E(t)- \tilde L^{down}_k\delta \theta_E + \delta P_H(t).
    \label{17}
\end{equation}
Defining, Hodge Laplacian using up and down Laplacian as $\tilde L_k = \tilde L^{up}_k + \tilde L^{down}_k$ and noting that $\tilde L_k \xi_H = 0$, we can simply rewrite the above Eq.~\eqref{17} as 
\begin{equation}
    \dot\delta \theta = \delta P (t) - \tilde L\delta \theta.
    \label{eq18}
\end{equation}
We can expand  $\delta \theta$ over the eigen-states $\mathbf{{u}_{\alpha}}$ with eigenvalues $\lambda_\alpha$ of $\tilde L_k$ as $\delta \theta(t) = \sum_\alpha c_\alpha(t) \mathbf{{u}_{\alpha}}$.  Thus we can rewrite \eqref{eq18} in terms of eigenmodes as
\begin{equation}
    \dot c_\alpha(t) = \delta P (t).\mathbf{{u}_{\alpha}}- \lambda_\alpha c_\alpha(t),
\end{equation}
where $\alpha = \{1,..,N_k\}$ with general solution
\begin{equation}
c_{\alpha}(t)
= e^{-\lambda_{\alpha} t} \, c_{\alpha}(0)
+ e^{-\lambda_{\alpha} t}
\int_{0}^{t} dt'\, e^{\lambda_{\alpha} t'} \, \delta {P}(t') \cdot \mathbf{{u}_{\alpha}},
\label{casolution}
\end{equation}
which is same as derived in the case of node Kuramoto model \cite{PhysRevLett.120.084101}.

\section{Deviation Measure}\label{sec5}
We can define an instantaneous measure of deviation of phases $\theta$ with time as
\begin{equation}
    \mathcal{D}(t) = |\theta(t) - \Delta(t) |^2,
\end{equation}
where $\Delta(t) = N_{k}^{-1}\sum_j^{N_k}  \theta_{j}(t)$.
\subsection{Analytic Calculation of $\mathcal{D}$}
We have defined, $\theta(t) = \theta^\ast_C + \delta \theta_C(t) + \theta^\ast_E + \delta \theta_E(t) + \xi_H(t) + \theta^\ast_{H,0} + \omega_Ht$. Equivalently in eigenmodes, we can write this as $ \theta(t) = \sum_\alpha c_\alpha(t) u_\alpha + \omega_Ht +  \theta^\ast  $, where we can define constant phase offsets as $\theta^\ast = \theta^\ast_C + \theta^\ast_E + \theta^\ast_{H,0} $. Now, 
\begin{equation}
    \Delta(t) = \frac{1}{n_1}\sum_{\alpha,j}c_{\alpha}(t) u_{\alpha,j} + \frac{1}{n_1}\sum_j \omega_{H,j} t + \frac{1}{n_1}\sum_j \theta_j^\ast 
\end{equation}
Lets denote $n_1^{-1} \sum_j \theta^\ast_j = \bar\theta^\ast$, $n_1^{-1}\sum_j \omega_{H,j} = \bar \omega_H $  and note that since eigenvectors are orthogonal so $\sum_j u_{\alpha,j} = 0$ for $\alpha \geq 2$. The projection over first eigenmode can be written as $\sum_j c_1(t)u_{1,j} = c_1(t)\sqrt{n_1}$ (since $u_1 = \dfrac{1}{\sqrt{n_1}}\mathbf{1}$).\\
So, $\Delta(t) =  c_1(t)u_1 +\bar \omega_H t  + \bar\theta^\ast$.
This gives 
\begin{equation}
    \mathcal{D}(t) = |\theta(t) - \Delta(t) |^2 = |\sum_{\alpha \geq 2} c_{\alpha}u_{\alpha} + (\omega_H - \bar\omega_H \mathbf{1})t +  (\theta^\ast - \bar\theta^\ast \mathbf{1}) |^2
    \label{D(t)}
\end{equation}
The time dependent harmonic component and constant offset have non-perturbative origin and doesn't depend on perturbation and its duration. Thus, the projected coefficients $c_\alpha (t)$ truly captures the response of perturbation. We can set $\omega_H = 0$ and initialize at $\theta^\ast = 0$ to get the spectral dependency of the perturbation similar to results obtained for the case of standard Kuramoto model \cite{PhysRevLett.120.084101}.
\\

\subsection{Measure of Fragility for a simplicial complex}
We can define a time integrated $\mathcal{D}(t)$ as a measure for fragility of a particular simplicial complex as
\begin{equation}
    \mathcal{C} = \int_0^T \mathcal{D}(t)dt.
\end{equation}
 
We will consider two types of perturbations, namely box and periodic perturbations.

\subsection{Box Perturbation} \label{box_sec}
First we consider a static perturbation $\delta P(t) = \delta P_0 \Theta(t_0) \Theta(\tau_0+t_0-t)$, where $\delta P_0$ is a vector representing the edges we choose to perturb. Thus, Eq.~\eqref{casolution} gives

\begin{equation}
c_{\alpha}(t)=
\begin{cases}
0,
& t\leq t_0,
\\[0.6em]

e^{-\lambda_\alpha t}c_\alpha(0)
+A_\alpha\left(1-e^{-\lambda_\alpha t}\right),
& t_0<t<t_0+\tau_0,
\\[0.8em]

e^{-\lambda_\alpha t}c_\alpha(0)
+A_\alpha
\left[
e^{\lambda_\alpha(\tau_0-t)}
-e^{-\lambda_\alpha t}
\right],
& t\geq t_0+\tau_0.
\end{cases}
\label{eq:calpha_solution}
\end{equation}

where $A_\alpha = \delta P_{0}\cdot \mathbf{u}_{\alpha}/\lambda_{\alpha}.$

\begin{figure}[htbp]
\centering
\begin{subfigure}{0.7\linewidth}
    \includegraphics[width=\linewidth]{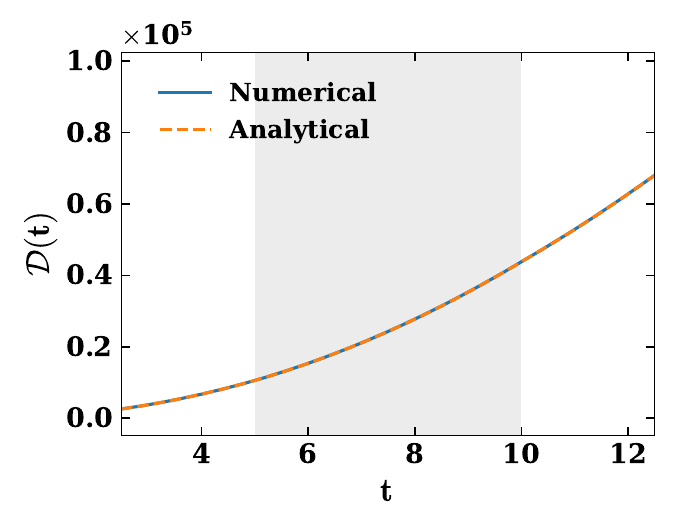}
    \caption{}
\end{subfigure}

\vspace{0.5em}
\begin{subfigure}{0.7\linewidth}
     \includegraphics[width=\linewidth]{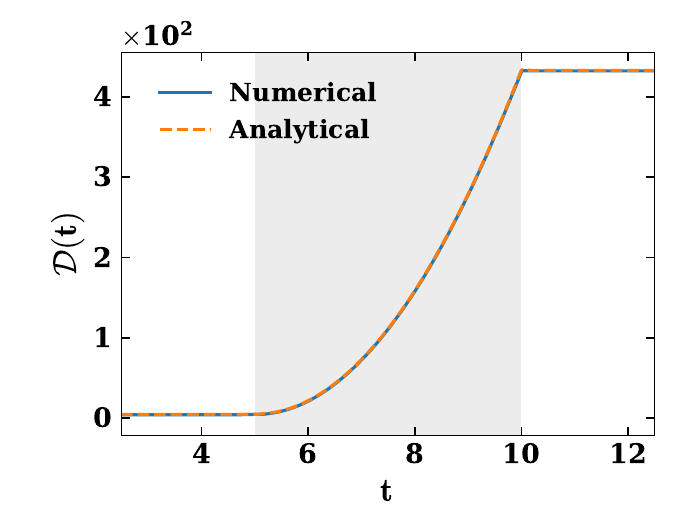}
    \caption{}
\end{subfigure}

\caption{For box perturbation on simplicial complex generated from configuration model with power-law exponent $\gamma=2.5$ and $m=2$ having $N_0=200, N_1 =866$ and $N_2=331$. We have applied perturbation for $\tau_0 = 5$ on 50 randomly selected edges with $P_0 = 1.0$. The grey region correspond to perturbation period. (a) We consider $\omega \sim \mathbf{U}(0,2)$ and harmonic components which are drifting dominate the evolution of phases. (b) We remove harmonic natural frequencies from $\omega$, i.e., $ \omega - \omega_H$ and see  the perturbation response.}
\label{fig:analyticalvsnum}
\end{figure}

For eigenmodes $\mathbf{u}_{\alpha}$ corresponding to $\lambda_{\alpha} =0$, $c_{\alpha}(t)$ reduces to $(\delta P_0 \cdot \mathbf{u}_{\alpha})t$ for $t_0 < t < t_0 + \tau_0$ and 0 otherwise. So, after the perturbation interval, $t> t_0 + \tau_0$, eigenmodes pertaining to $\lambda_\alpha = 0$ do not contribute to $\mathcal{D}(t)$.  This signifies the non-decaying flows pertaining to harmonic modes during the perturbation interval. Since, Hodge subspaces are orthogonal,  $\delta P_0 \cdot \mathbf{u}_{\alpha} = \delta P_H \cdot \mathbf{u}_{\alpha}$, where $\delta P_H$ is harmonic component of perturbation $\delta P$.  In Fig.~\ref{fig:analyticalvsnum} we have plotted the analytical and numerical values for $\mathcal{D}(t)$ for perturbation applied on 50 randomly selected edges for cases where we have included and excluded harmonic natural frequencies. The drifting harmonic component adds to the deviation measure thus we can take $\omega_H = 0$ to isolate the perturbation response.

We can introduce perturbation at $t_0 = 0$ without loss of generality, thus can set  $c_{\alpha}(0) = 0$. 
The fragility measure in this case for $T > \tau_0$ with $\omega_H = \delta P_H=0$ and neglecting the offsets reduces $\mathcal{C}$ to $\int_0^T c_\alpha^2(t)dt$ which in case of box perturbation is given as 

\begin{equation}
    \mathcal{C} = \sum_{\alpha > \beta_1} \frac{(\delta P_0 \cdot \mathbf{u}_\alpha)^2}{\lambda_\alpha^3} \left( \lambda_\alpha \tau_0 - 1 + e^{-\lambda_\alpha \tau_0} \right).
\end{equation}
 This result coincide with study of box perturbation on the node based Kuramoto model \cite{PhysRevLett.120.084101}.
For $c_\alpha$ dominating $\mathcal{D}(t)$, for which we have to take $\omega_H=0$ and neglecting fixed offset, $\mathcal{C}$  depends upon generalized Kirchhoff indices $(Kf_m)$ \cite{Zhu1996,Gutman1996} summed over non-zero eigenvalues
 \begin{equation}
     \mathcal{C} \propto \sum_{\alpha>\beta_1}\lambda^{-m}_\alpha.
 \end{equation}

\begin{figure}[htbp]
\centering
\begin{subfigure}{0.7\linewidth}
    \includegraphics[width=\linewidth]{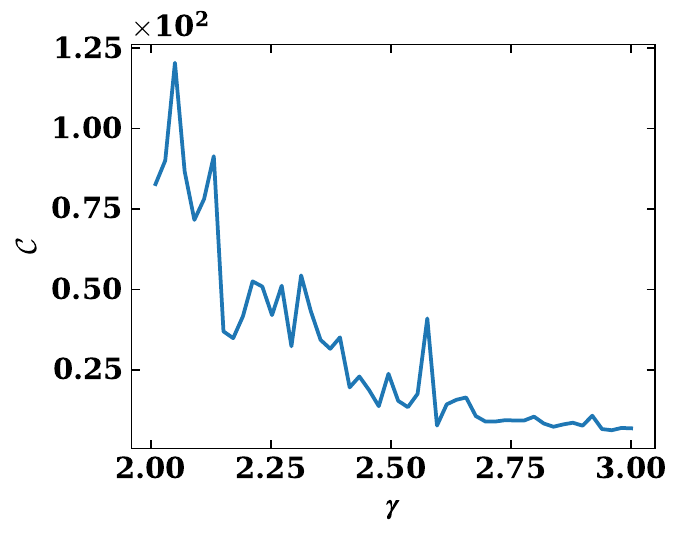}
    \caption{}
\end{subfigure}

    \vspace{0.5em}
    
    \begin{subfigure}{0.7\linewidth}
        \includegraphics[width=\linewidth]{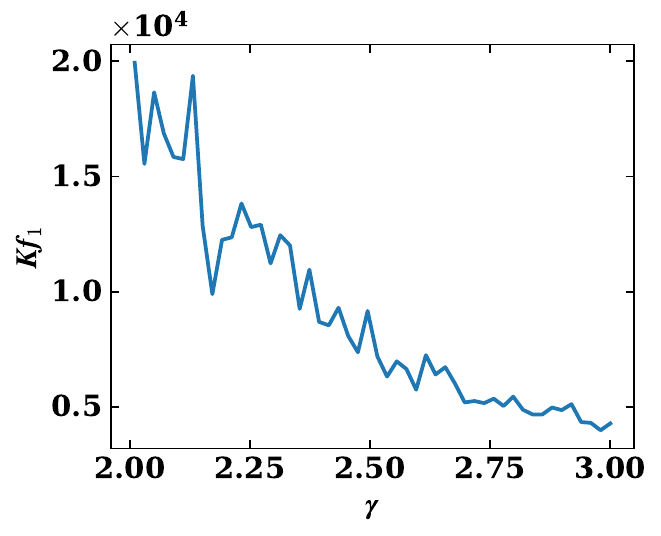}
        \caption{}
    \end{subfigure}

\caption{Robustness of a simplicial complex under box perturbation for 10 averaged realization varying $\gamma$ for $m=2$ having $N_0 = 100$ and  neglecting the Harmonic contribution by taking harmonic natural frequency and harmonic perturbation component to be zero. The perturbation is applied on 20 randomly selected edges.}
\label{fig:comparisonboxkf1vsC}
\end{figure}

We  plot $\mathcal{C}$ by varying $\gamma$ for simplicial configuration model for box perturbation  in Fig.~\ref{fig:comparisonboxkf1vsC} with $\omega_H = \delta P_H = 0$ as harmonic component of perturbation induce persisting flows that accumulates in $\mathcal{C}$ and depends upon perturbation interval and compared with $Kf_1 = \sum_{\alpha > \beta_1} \lambda_\alpha^{-1}$. The generalized Kirchhoff indices, $Kf_m$ carry same qualitative trend as $\lambda_\alpha > 0$ while being sensitive to small eigenvalues.

\begin{figure}[htbp]
\centering

\begin{subfigure}{0.7\linewidth}
    \includegraphics[width=\linewidth]{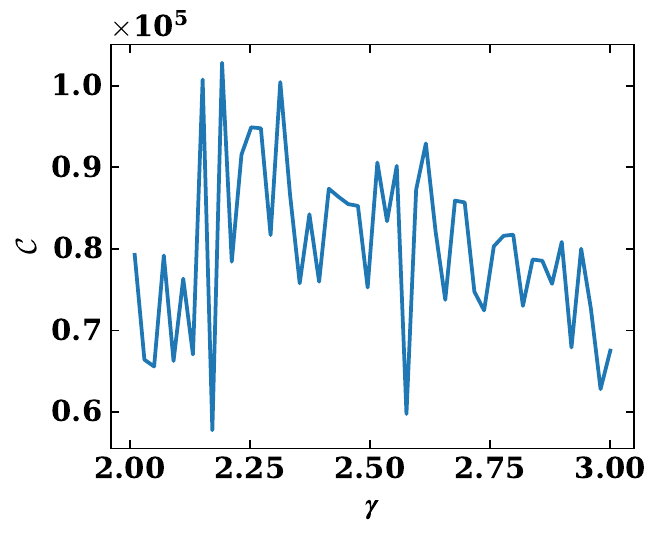}
    \caption{}
\end{subfigure}

\vspace{0.5em}
\begin{subfigure}{0.7\linewidth}
    \includegraphics[width=\linewidth]{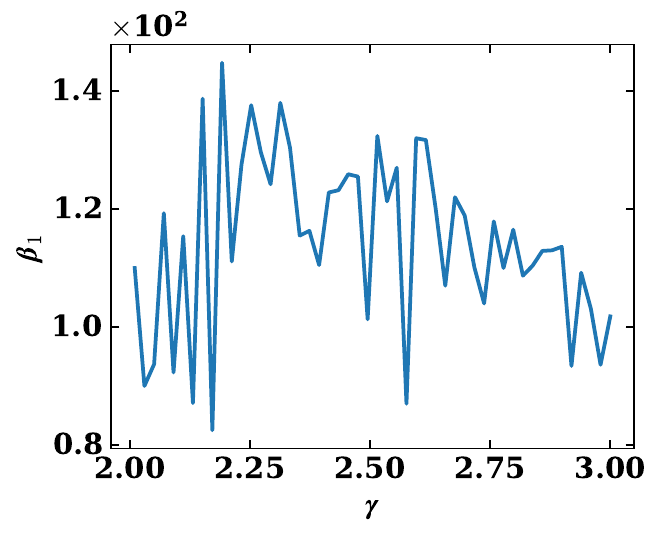}
    \caption{}
\end{subfigure}

\caption{Robustness of a simplicial complex  for 10 averaged realization
varying $\gamma$ for $m=2$ including the Harmonic contribution. (a),(b) suggests the deviation measure dominated by the harmonic component of phases and is proportional to $\beta_1$.}
\label{fig:comparisonbox_Cvsbeta1}
\end{figure}

Inclusion of Harmonic component makes $(\omega_H-\bar\omega_H\mathbf{1})t$ term dominate  $\mathcal{D}(t)$ in Eq.~\eqref{D(t)}.
Thus as, the dimension of harmonic space grows, $\mathcal{D}(t)$ is primarily driven by harmonic component. Therefore, inclusion of Harmonic component makes
 \begin{equation}
     \mathcal{C} \propto \beta_1
 \end{equation}
To verify this we compare $\mathcal{C}$ with $\beta_1$ by varying $\gamma$ for simplicial configuration model in Fig.~\ref{fig:comparisonbox_Cvsbeta1}.

\subsection{Periodic Perturbation }
We can next analyze periodic perturbation on reference state of the system introduced at $t = 0$. The real world perturbations like those in  electric grids, neurons and biological rhythms are approximately periodic. Therefore, we consider $\delta P(t) = \delta P_0 \sin(\Omega t)\Theta(t) \Theta(\tau_0-t)$, where $\delta P_0$ is a vector containing the perturbation strength for each selected edge.
The corresponding projection coefficient is given by
\begin{equation}
    c_{\alpha}(t)= \left\{
\begin{array}{@{}l@{\;}l@{}}
  A_{\alpha}[\lambda_\alpha \sin(\Omega t) - \Omega \cos(\Omega t) + \Omega e^{-\lambda_\alpha t}]
& t \le \tau_{0}, \\[0.8em]
\begin{aligned}
    & A_\alpha[\lambda_\alpha e^{\lambda_\alpha (\tau_0 -t)}\sin(\Omega \tau_0) \\[-0.2em]
&\quad{}- \Omega e^{\lambda_\alpha (\tau_0 -t)}\cos(\Omega \tau_0) + \Omega e^{-\lambda_\alpha t}],
\end{aligned}
& t > \tau_{0}.
\end{array} \right.
\end{equation}
where $A_\alpha = \dfrac{\delta P_{0}\cdot \mathbf{u}_{\alpha}}{\lambda_\alpha^2 + \Omega^2} $
For eigenmodes $\mathbf{u}_\alpha$ corresponding to $\lambda_\alpha = 0$, $c_\alpha$ reduces to $\dfrac{\delta P \cdot \mathbf{u}_\alpha}{\Omega}(1 - \cos(\Omega t))$ for $t \leq \tau_0$ signifying a bounded oscillation of harmonic modes during perturbation interval unlike the case of box perturbation which shows linear drift. For $t> \tau_0$, it freezes at permanent offset $\frac{\delta P \cdot \mathbf{u}_\alpha}{\Omega}(1 - \cos(\Omega \tau_0))$. If $\Omega \tau_0 \neq 2 \pi n , n \in \mathbb{Z}$, it leaves a trace of perturbation in the harmonic sector. 

\begin{figure}[htbp]
\centering
\begin{subfigure}{0.7\linewidth}
    \includegraphics[width=\linewidth]{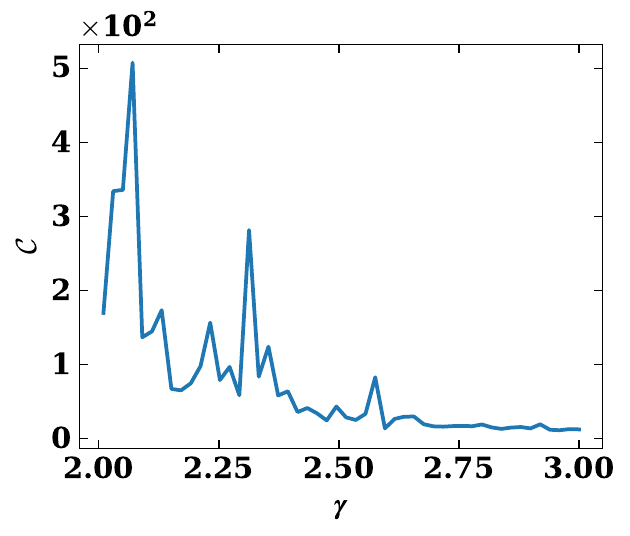}
    \caption{}
\end{subfigure}
 
\vspace{0.5em}

\begin{subfigure}{0.7\linewidth}
    \includegraphics[width=\linewidth]{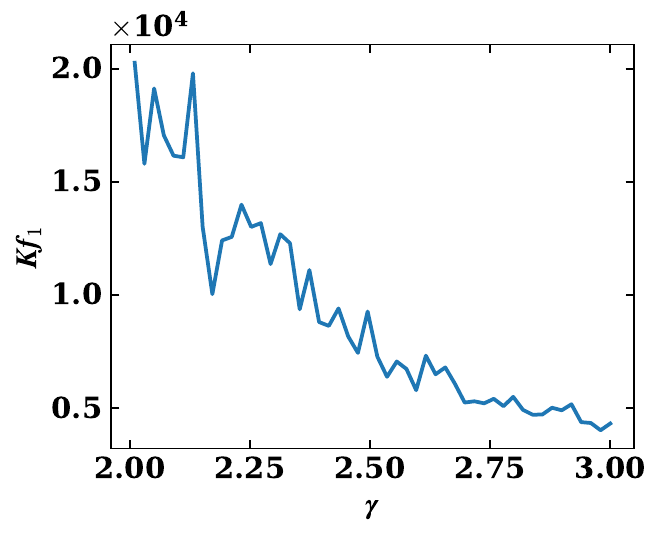}
    \caption{}
\end{subfigure}

\caption{Robustness of a simplicial complex  under periodic perturbation having $P_0 = 1, \Omega=1$ for 10 averaged realization varying $\gamma$ for $m=2$ having $N_0 = 100$ and neglecting the Harmonic contribution taking harmonic natural frequency and harmonic perturbation component to be zero.}
\label{fig:comparisonboxkf1vsCperiodic}
\end{figure}

We can take $\omega_H = \delta P_H=0$ and neglect fixed offsets which reduces $\mathcal{D}(t)$ to $c_\alpha^2(t)$. Therefore, taking long perturbation $\lambda_\alpha \tau_0 \ll 1, \hspace{0.01cm} \forall \alpha $ we get $\mathcal{C}$ for periodic perturbation as
\begin{equation}
    \mathcal{C} \sim \dfrac{\tau_0}{2}\sum_{\alpha > \beta_1} \dfrac{\delta P_{0}\cdot \mathbf{u}_{\alpha}}{\lambda_\alpha^2 + \Omega^2}.
\end{equation}
We compare $\mathcal{C}$ with $Kf_1$ in Fig.~\ref{fig:comparisonboxkf1vsCperiodic} by varying $\gamma$ for simplicial configuration model for $\Omega =1$ and $P_0 = 1$, applied on 20 randomly selected edges.

\section{Role of Topology}\label{sec6}
\begin{figure}[htbp]
    \centering

    \begin{subfigure}[b]{0.48\columnwidth}
        \centering
        \includegraphics[width=\linewidth]{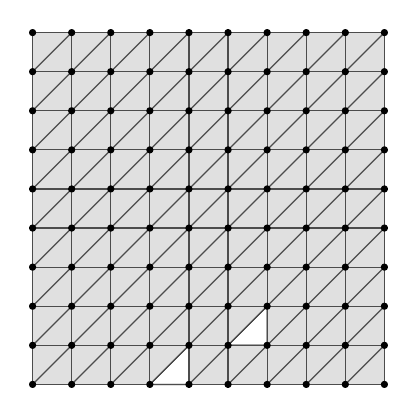}
        \caption{}
    \end{subfigure}
    \hfill
    \begin{subfigure}[b]{0.48\columnwidth}
        \centering
        \includegraphics[width=\linewidth]{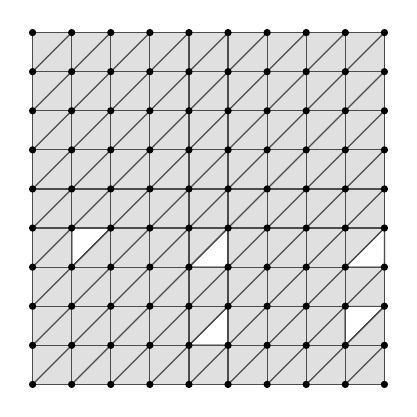}
        \caption{}
    \end{subfigure}
\caption{Triangulated torus with periodic boundary condition. (a) two hollowed triangles giving $\beta_1 = 2$, (b) five triangles not taken into consideration giving $\beta_1 = 5$.}
\label{toriconstruction}
    
\end{figure}

From Sec.~\ref{box_sec} we see that robustness depends upon cavities, i.e., for dynamics defined on edges it depends upon first Betti number, $\beta_1$ as the harmonic flows persist during the perturbation interval.  To see how topology affects the perturbation response, we take a triangulated torus with desired number of 2-simplex thus allowing control over $\beta_1$ as illustrated in Fig.~\ref{toriconstruction}.

\begin{figure}[htbp]
    \centering

    \begin{subfigure}{0.7\linewidth}
        \centering
        \includegraphics[width=\linewidth]{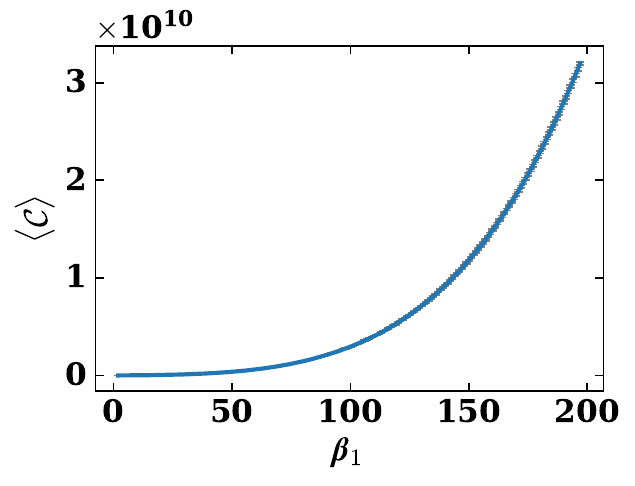}
        \caption{}
        \label{fig:tori_beta1}
    \end{subfigure}

    \vspace{0.5em}

    \begin{subfigure}{0.7\linewidth}
        \centering
        \includegraphics[width=\linewidth]{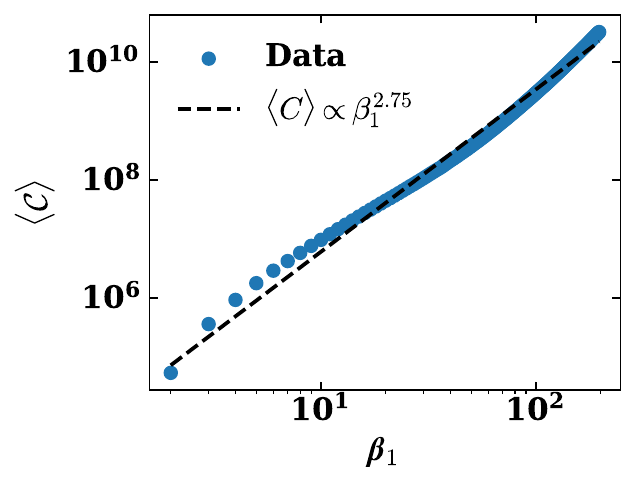}
        \caption{}
        \label{fig:robustness_scaling}
    \end{subfigure}

    \caption{Perturbation response of triangulated tori with 100 nodes averaged over 10 realization with desired number of $\beta_1$ for box perturbation applied on 20 randomly chosen edges  for each realization with $\delta P_0 =1$. The response (a), shows positive correlation with $\beta_1$. (b). $\langle\mathcal{C}\rangle$ shows scaling exponent 2.75 with $R^2 = 0.99$.}
    \label{fig:scaling}
\end{figure}

The dynamics is defined on the edges of the triangulated tori. 50 edges are randomly selected and subjected to box perturbation with $P_0 = 1$. To  measure the effect of perturbation, we remove the harmonic natural frequency i.e., $\omega_H =0$ from the initialization of $\omega \sim \mathbf{U}(1,3)$. With this setup, we plot $\mathcal{C}$ averaged over 10 realizations with $\beta_1$ in Fig.~\ref{fig:scaling}.

We observe that $\langle \mathcal{C} \rangle$ exhibits power law scaling with $\beta_1$ with an exponent of 2.75.  
The harmonic part of perturbation $P_H$ causes persistent oscillation in harmonic sector as its non-dissipative. The coexact and exact sectors being dissipative and therefore, damps their respective perturbation component. Thus, an increase in dimension of harmonic subspace leads to persistent oscillation thus the measure $\mathcal{C}$ increases with $\beta_1$.  

\section{Conclusion}\label{sec7}
We have studied the Hodge-decomposed simplicial Kuramoto  dynamics  defined on dimension, $k$. The dynamics on decomposed Hodge subspaces evolve independently allowing us to study them separately. Synchronization in the classical sense does not extend cleanly to $k \geq 1$ as the harmonic section of dimension $\beta_k$ drifts freely with its own natural frequency. For $\beta_k > 1$, the harmonic sector supports multiple independent drifting directions and the uncompounded phase vector cannot achieve a fully synchronized state. This obstruction is not an artifact of the sinusoidal Kuramoto coupling and follows from orthogonality relations of Hodge decomposition (Eqs.~\eqref{orth1},\eqref{orth2}), which holds for any dynamics whose interaction term is mediated through the boundary operators $B_k$ and $B_{k+1}$.

Despite this obstruction, we show that the coexact and exact sectors admit a stable fixed point above an explicitly derived, topology-dependent critical coupling strength (Eqs.~\eqref{coexact critical},\eqref{exact critical}). 

The synchronization of classical Kuramoto model (Sec.~\ref{sec33}) can be understood as coexact component goes into fixed phase state and harmonic component which is one dimensional for connected graph rotates with mean of natural frequency. Hodge decomposed dynamics defined on $k\geq 1$ show similar feature, as self interacting subspaces, coexact and exact goes into fixed phase state beyond the critical coupling strength. Thus, Hodge decomposed dynamics show a consistent picture of synchronization across dimensions where non-harmonic sectors show fixed phase state.

We then take this fixed point state as reference to study perturbation response of the dynamics defined on simplices of dimension $k$. We again take the Hodge theoretic route and decomposed the perturbation itself leading to autonomous subspace evolution. The perturbation response is governed by the dynamical Eq.~\eqref{eq18}. To quantify the perturbation response we introduced a deviation measure, $\mathcal{D}$ (Eq.~\eqref{D(t)}). The measure is contaminated by harmonic modes so we set $\omega_H$ to 0 and observe that for non-harmonic perturbation fragility measure $\mathcal{C}$ depends on generalized Kirchhoff indices. Retaining the harmonic natural frequency and perturbation component, we see that as the dimension of harmonic subspace grows, oscillations persist thus $\mathcal{C}$ depends upon $\beta_k$.
The coexact and exact sectors are dissipative and damps the perturbation while the harmonic phases drifts without any interaction. 
We analyze this for two  types of perturbation, box perturbation and periodic perturbation. 
Finally, to see the role of topology in the perturbation response of edge based Kuramoto model,  we study box perturbation on triangulated torus with desired $\beta_1$ and observe that for $\omega_H=0$, fragility scales superlinearly with $\beta_1$ as perturbations persists in those harmonic modes. 

An important direction this raises is what happens to the topologically obstructed harmonic sector under more general higher-order coupling. In the present model, the sinusoidal coupling's algebraic structure keeps the harmonic sector fully decoupled from the exact and coexact dynamics, so the topological obstruction manifests purely as free drift. For a general nonlinear coupling, this need not hold, the harmonic modes while still invisible to the boundary operators individually could become coupled to exact and coexact sectors producing frustration rather than clean decoupling. Whether such topologically induced frustration is a generic phenomenon across broader classes of dynamics on simplicial complexes and how it would reshape the fragility scaling derived here is a natural and structurally motivated question for future study. 

\section*{DATA AVAILABILITY}
The data that support the findings of this study are
available from the corresponding author upon reasonable
request.



\begin{thebibliography}{40}%
	\makeatletter
	\providecommand \@ifxundefined [1]{%
		\@ifx{#1\undefined}
	}%
	\providecommand \@ifnum [1]{%
		\ifnum #1\expandafter \@firstoftwo
		\else \expandafter \@secondoftwo
		\fi
	}%
	\providecommand \@ifx [1]{%
		\ifx #1\expandafter \@firstoftwo
		\else \expandafter \@secondoftwo
		\fi
	}%
	\providecommand \natexlab [1]{#1}%
	\providecommand \enquote  [1]{``#1''}%
	\providecommand \bibnamefont  [1]{#1}%
	\providecommand \bibfnamefont [1]{#1}%
	\providecommand \citenamefont [1]{#1}%
	\providecommand \href@noop [0]{\@secondoftwo}%
	\providecommand \href [0]{\begingroup \@sanitize@url \@href}%
	\providecommand \@href[1]{\@@startlink{#1}\@@href}%
	\providecommand \@@href[1]{\endgroup#1\@@endlink}%
	\providecommand \@sanitize@url [0]{\catcode `\\12\catcode `\$12\catcode
		`\&12\catcode `\#12\catcode `\^12\catcode `\_12\catcode `\%12\relax}%
	\providecommand \@@startlink[1]{}%
	\providecommand \@@endlink[0]{}%
	\providecommand \url  [0]{\begingroup\@sanitize@url \@url }%
	\providecommand \@url [1]{\endgroup\@href {#1}{\urlprefix }}%
	\providecommand \urlprefix  [0]{URL }%
	\providecommand \Eprint [0]{\href }%
	\providecommand \doibase [0]{https://doi.org/}%
	\providecommand \selectlanguage [0]{\@gobble}%
	\providecommand \bibinfo  [0]{\@secondoftwo}%
	\providecommand \bibfield  [0]{\@secondoftwo}%
	\providecommand \translation [1]{[#1]}%
	\providecommand \BibitemOpen [0]{}%
	\providecommand \bibitemStop [0]{}%
	\providecommand \bibitemNoStop [0]{.\EOS\space}%
	\providecommand \EOS [0]{\spacefactor3000\relax}%
	\providecommand \BibitemShut  [1]{\csname bibitem#1\endcsname}%
	\let\auto@bib@innerbib\@empty
	\bibitem [{\citenamefont {Kuramoto}(1975)}]{10.1007/BFb0013365}%
	\BibitemOpen
	\bibfield  {author} {\bibinfo {author} {\bibfnamefont {Y.}~\bibnamefont
			{Kuramoto}},\ }\bibfield  {title} {\bibinfo {title} {Self-entrainment of a
			population of coupled non-linear oscillators},\ }in\ \href@noop {} {\emph
		{\bibinfo {booktitle} {International Symposium on Mathematical Problems in
				Theoretical Physics}}},\ \bibinfo {editor} {edited by\ \bibinfo {editor}
		{\bibfnamefont {H.}~\bibnamefont {Araki}}}\ (\bibinfo  {publisher} {Springer
		Berlin Heidelberg},\ \bibinfo {address} {Berlin, Heidelberg},\ \bibinfo
	{year} {1975})\ pp.\ \bibinfo {pages} {420--422}\BibitemShut {NoStop}%
	\bibitem [{\citenamefont {Acebr{\'o}n}\ \emph {et~al.}(2005)\citenamefont
		{Acebr{\'o}n}, \citenamefont {P{\'e}rez~Vicente}, \citenamefont {Ritort},\
		and\ \citenamefont {Spigler}}]{acebron2005kuramoto}%
	\BibitemOpen
	\bibfield  {author} {\bibinfo {author} {\bibfnamefont {J.~A.}\ \bibnamefont
			{Acebr{\'o}n}}, \bibinfo {author} {\bibfnamefont {C.~J.}\ \bibnamefont
			{P{\'e}rez~Vicente}}, \bibinfo {author} {\bibfnamefont {F.}~\bibnamefont
			{Ritort}},\ and\ \bibinfo {author} {\bibfnamefont {R.}~\bibnamefont
			{Spigler}},\ }\bibfield  {title} {\bibinfo {title} {The kuramoto model: A
			simple paradigm for synchronization phenomena},\ }\href@noop {} {\bibfield
		{journal} {\bibinfo  {journal} {Reviews of modern physics}\ }\textbf
		{\bibinfo {volume} {77}},\ \bibinfo {pages} {137} (\bibinfo {year}
		{2005})}\BibitemShut {NoStop}%
	\bibitem [{\citenamefont {Breakspear}\ \emph {et~al.}(2010)\citenamefont
		{Breakspear}, \citenamefont {Heitmann},\ and\ \citenamefont
		{Daffertshofer}}]{10.3389/fnhum.2010.00190}%
	\BibitemOpen
	\bibfield  {author} {\bibinfo {author} {\bibfnamefont {M.}~\bibnamefont
			{Breakspear}}, \bibinfo {author} {\bibfnamefont {S.}~\bibnamefont
			{Heitmann}},\ and\ \bibinfo {author} {\bibfnamefont {A.}~\bibnamefont
			{Daffertshofer}},\ }\bibfield  {title} {\bibinfo {title} {Generative models
			of cortical oscillations: Neurobiological implications of the kuramoto
			model},\ }\href@noop {} {\bibfield  {journal} {\bibinfo  {journal} {Frontiers
				in Human Neuroscience}\ }\textbf {\bibinfo {volume} {Volume 4 - 2010}}
		(\bibinfo {year} {2010})}\BibitemShut {NoStop}%
	\bibitem [{\citenamefont {Strogatz}(1997)}]{638513}%
	\BibitemOpen
	\bibfield  {author} {\bibinfo {author} {\bibfnamefont {S.}~\bibnamefont
			{Strogatz}},\ }\bibfield  {title} {\bibinfo {title} {Spontaneous
			synchronization in nature},\ }in\ \href@noop {} {\emph {\bibinfo {booktitle}
			{Proceedings of International Frequency Control Symposium}}}\ (\bibinfo
	{year} {1997})\ pp.\ \bibinfo {pages} {2--4}\BibitemShut {NoStop}%
	\bibitem [{\citenamefont {N\'eda}\ \emph {et~al.}(2000)\citenamefont {N\'eda},
		\citenamefont {Ravasz}, \citenamefont {Vicsek}, \citenamefont {Brechet},\
		and\ \citenamefont {Barab\'asi}}]{PhysRevE.61.6987}%
	\BibitemOpen
	\bibfield  {author} {\bibinfo {author} {\bibfnamefont {Z.}~\bibnamefont
			{N\'eda}}, \bibinfo {author} {\bibfnamefont {E.}~\bibnamefont {Ravasz}},
		\bibinfo {author} {\bibfnamefont {T.}~\bibnamefont {Vicsek}}, \bibinfo
		{author} {\bibfnamefont {Y.}~\bibnamefont {Brechet}},\ and\ \bibinfo {author}
		{\bibfnamefont {A.~L.}\ \bibnamefont {Barab\'asi}},\ }\bibfield  {title}
	{\bibinfo {title} {Physics of the rhythmic applause},\ }\href@noop {}
	{\bibfield  {journal} {\bibinfo  {journal} {Phys. Rev. E}\ }\textbf {\bibinfo
			{volume} {61}},\ \bibinfo {pages} {6987} (\bibinfo {year}
		{2000})}\BibitemShut {NoStop}%
	\bibitem [{\citenamefont {Rohden}\ \emph {et~al.}(2012)\citenamefont {Rohden},
		\citenamefont {Sorge}, \citenamefont {Timme},\ and\ \citenamefont
		{Witthaut}}]{PhysRevLett.109.064101}%
	\BibitemOpen
	\bibfield  {author} {\bibinfo {author} {\bibfnamefont {M.}~\bibnamefont
			{Rohden}}, \bibinfo {author} {\bibfnamefont {A.}~\bibnamefont {Sorge}},
		\bibinfo {author} {\bibfnamefont {M.}~\bibnamefont {Timme}},\ and\ \bibinfo
		{author} {\bibfnamefont {D.}~\bibnamefont {Witthaut}},\ }\bibfield  {title}
	{\bibinfo {title} {Self-organized synchronization in decentralized power
			grids},\ }\href@noop {} {\bibfield  {journal} {\bibinfo  {journal} {Phys.
				Rev. Lett.}\ }\textbf {\bibinfo {volume} {109}},\ \bibinfo {pages} {064101}
		(\bibinfo {year} {2012})}\BibitemShut {NoStop}%
	\bibitem [{\citenamefont {Barahona}\ and\ \citenamefont
		{Pecora}(2002)}]{PhysRevLett.89.054101}%
	\BibitemOpen
	\bibfield  {author} {\bibinfo {author} {\bibfnamefont {M.}~\bibnamefont
			{Barahona}}\ and\ \bibinfo {author} {\bibfnamefont {L.~M.}\ \bibnamefont
			{Pecora}},\ }\bibfield  {title} {\bibinfo {title} {Synchronization in
			small-world systems},\ }\href@noop {} {\bibfield  {journal} {\bibinfo
			{journal} {Phys. Rev. Lett.}\ }\textbf {\bibinfo {volume} {89}},\ \bibinfo
		{pages} {054101} (\bibinfo {year} {2002})}\BibitemShut {NoStop}%
	\bibitem [{\citenamefont {Iacopini}\ \emph {et~al.}(2019)\citenamefont
		{Iacopini}, \citenamefont {Petri}, \citenamefont {Barrat},\ and\
		\citenamefont {Latora}}]{Iacopini2019}%
	\BibitemOpen
	\bibfield  {author} {\bibinfo {author} {\bibfnamefont {I.}~\bibnamefont
			{Iacopini}}, \bibinfo {author} {\bibfnamefont {G.}~\bibnamefont {Petri}},
		\bibinfo {author} {\bibfnamefont {A.}~\bibnamefont {Barrat}},\ and\ \bibinfo
		{author} {\bibfnamefont {V.}~\bibnamefont {Latora}},\ }\bibfield  {title}
	{\bibinfo {title} {Simplicial models of social contagion},\ }\href@noop {}
	{\bibfield  {journal} {\bibinfo  {journal} {Nature Communications}\ }\textbf
		{\bibinfo {volume} {10}},\ \bibinfo {pages} {2485} (\bibinfo {year}
		{2019})}\BibitemShut {NoStop}%
	\bibitem [{\citenamefont {Petri}\ and\ \citenamefont
		{Barrat}(2018)}]{PhysRevLett.121.228301}%
	\BibitemOpen
	\bibfield  {author} {\bibinfo {author} {\bibfnamefont {G.}~\bibnamefont
			{Petri}}\ and\ \bibinfo {author} {\bibfnamefont {A.}~\bibnamefont {Barrat}},\
	}\bibfield  {title} {\bibinfo {title} {Simplicial activity driven model},\
	}\href@noop {} {\bibfield  {journal} {\bibinfo  {journal} {Phys. Rev. Lett.}\
		}\textbf {\bibinfo {volume} {121}},\ \bibinfo {pages} {228301} (\bibinfo
		{year} {2018})}\BibitemShut {NoStop}%
	\bibitem [{\citenamefont {Boccaletti}\ \emph {et~al.}(2023)\citenamefont
		{Boccaletti}, \citenamefont {{De Lellis}}, \citenamefont {{del Genio}},
		\citenamefont {Alfaro-Bittner}, \citenamefont {Criado}, \citenamefont
		{Jalan},\ and\ \citenamefont {Romance}}]{BOCCALETTI20231}%
	\BibitemOpen
	\bibfield  {author} {\bibinfo {author} {\bibfnamefont {S.}~\bibnamefont
			{Boccaletti}}, \bibinfo {author} {\bibfnamefont {P.}~\bibnamefont {{De
					Lellis}}}, \bibinfo {author} {\bibfnamefont {C.}~\bibnamefont {{del Genio}}},
		\bibinfo {author} {\bibfnamefont {K.}~\bibnamefont {Alfaro-Bittner}},
		\bibinfo {author} {\bibfnamefont {R.}~\bibnamefont {Criado}}, \bibinfo
		{author} {\bibfnamefont {S.}~\bibnamefont {Jalan}},\ and\ \bibinfo {author}
		{\bibfnamefont {M.}~\bibnamefont {Romance}},\ }\bibfield  {title} {\bibinfo
		{title} {The structure and dynamics of networks with higher order
			interactions},\ }\href@noop {} {\bibfield  {journal} {\bibinfo  {journal}
			{Physics Reports}\ }\textbf {\bibinfo {volume} {1018}},\ \bibinfo {pages} {1}
		(\bibinfo {year} {2023})},\ \bibinfo {note} {the structure and dynamics of
		networks with higher order interactions}\BibitemShut {NoStop}%
	\bibitem [{\citenamefont {Battiston}\ \emph {et~al.}(2020)\citenamefont
		{Battiston}, \citenamefont {Cencetti}, \citenamefont {Iacopini},
		\citenamefont {Latora}, \citenamefont {Lucas}, \citenamefont {Patania},
		\citenamefont {Young},\ and\ \citenamefont {Petri}}]{battiston2020networks}%
	\BibitemOpen
	\bibfield  {author} {\bibinfo {author} {\bibfnamefont {F.}~\bibnamefont
			{Battiston}}, \bibinfo {author} {\bibfnamefont {G.}~\bibnamefont {Cencetti}},
		\bibinfo {author} {\bibfnamefont {I.}~\bibnamefont {Iacopini}}, \bibinfo
		{author} {\bibfnamefont {V.}~\bibnamefont {Latora}}, \bibinfo {author}
		{\bibfnamefont {M.}~\bibnamefont {Lucas}}, \bibinfo {author} {\bibfnamefont
			{A.}~\bibnamefont {Patania}}, \bibinfo {author} {\bibfnamefont {J.-G.}\
			\bibnamefont {Young}},\ and\ \bibinfo {author} {\bibfnamefont
			{G.}~\bibnamefont {Petri}},\ }\bibfield  {title} {\bibinfo {title} {Networks
			beyond pairwise interactions: Structure and dynamics},\ }\href@noop {}
	{\bibfield  {journal} {\bibinfo  {journal} {Physics Reports}\ }\textbf
		{\bibinfo {volume} {874}},\ \bibinfo {pages} {1} (\bibinfo {year}
		{2020})}\BibitemShut {NoStop}%
	\bibitem [{\citenamefont {Bianconi}(2021)}]{bianconi2021higher}%
	\BibitemOpen
	\bibfield  {author} {\bibinfo {author} {\bibfnamefont {G.}~\bibnamefont
			{Bianconi}},\ }\href@noop {} {\emph {\bibinfo {title} {Higher-order
				networks}}}\ (\bibinfo  {publisher} {Cambridge University Press},\ \bibinfo
	{year} {2021})\BibitemShut {NoStop}%
	\bibitem [{\citenamefont {Jiang}\ \emph {et~al.}(2011)\citenamefont {Jiang},
		\citenamefont {Lim}, \citenamefont {Yao},\ and\ \citenamefont
		{Ye}}]{Jiang2011}%
	\BibitemOpen
	\bibfield  {author} {\bibinfo {author} {\bibfnamefont {X.}~\bibnamefont
			{Jiang}}, \bibinfo {author} {\bibfnamefont {L.-H.}\ \bibnamefont {Lim}},
		\bibinfo {author} {\bibfnamefont {Y.}~\bibnamefont {Yao}},\ and\ \bibinfo
		{author} {\bibfnamefont {Y.}~\bibnamefont {Ye}},\ }\bibfield  {title}
	{\bibinfo {title} {Statistical ranking and combinatorial hodge theory},\
	}\href@noop {} {\bibfield  {journal} {\bibinfo  {journal} {Mathematical
				Programming}\ }\textbf {\bibinfo {volume} {127}},\ \bibinfo {pages} {203}
		(\bibinfo {year} {2011})}\BibitemShut {NoStop}%
	\bibitem [{\citenamefont {Barbarossa}\ and\ \citenamefont
		{Sardellitti}(2020{\natexlab{a}})}]{9044758}%
	\BibitemOpen
	\bibfield  {author} {\bibinfo {author} {\bibfnamefont {S.}~\bibnamefont
			{Barbarossa}}\ and\ \bibinfo {author} {\bibfnamefont {S.}~\bibnamefont
			{Sardellitti}},\ }\bibfield  {title} {\bibinfo {title} {Topological signal
			processing over simplicial complexes},\ }\href@noop {} {\bibfield  {journal}
		{\bibinfo  {journal} {IEEE Transactions on Signal Processing}\ }\textbf
		{\bibinfo {volume} {68}},\ \bibinfo {pages} {2992} (\bibinfo {year}
		{2020}{\natexlab{a}})}\BibitemShut {NoStop}%
	\bibitem [{\citenamefont {Barbarossa}\ and\ \citenamefont
		{Sardellitti}(2020{\natexlab{b}})}]{barbarossa2020topological}%
	\BibitemOpen
	\bibfield  {author} {\bibinfo {author} {\bibfnamefont {S.}~\bibnamefont
			{Barbarossa}}\ and\ \bibinfo {author} {\bibfnamefont {S.}~\bibnamefont
			{Sardellitti}},\ }\bibfield  {title} {\bibinfo {title} {Topological signal
			processing over simplicial complexes},\ }\href@noop {} {\bibfield  {journal}
		{\bibinfo  {journal} {IEEE Transactions on Signal Processing}\ }\textbf
		{\bibinfo {volume} {68}},\ \bibinfo {pages} {2992} (\bibinfo {year}
		{2020}{\natexlab{b}})}\BibitemShut {NoStop}%
	\bibitem [{\citenamefont {Torres}\ and\ \citenamefont
		{Bianconi}(2020)}]{torres2020simplicial}%
	\BibitemOpen
	\bibfield  {author} {\bibinfo {author} {\bibfnamefont {J.~J.}\ \bibnamefont
			{Torres}}\ and\ \bibinfo {author} {\bibfnamefont {G.}~\bibnamefont
			{Bianconi}},\ }\bibfield  {title} {\bibinfo {title} {Simplicial complexes:
			higher-order spectral dimension and dynamics},\ }\href@noop {} {\bibfield
		{journal} {\bibinfo  {journal} {Journal of Physics: Complexity}\ }\textbf
		{\bibinfo {volume} {1}},\ \bibinfo {pages} {015002} (\bibinfo {year}
		{2020})}\BibitemShut {NoStop}%
	\bibitem [{\citenamefont {Tero}\ \emph {et~al.}(2010)\citenamefont {Tero},
		\citenamefont {Takagi}, \citenamefont {Saigusa}, \citenamefont {Ito},
		\citenamefont {Bebber}, \citenamefont {Fricker}, \citenamefont {Yumiki},
		\citenamefont {Kobayashi},\ and\ \citenamefont
		{Nakagaki}}]{doi:10.1126/science.1177894}%
	\BibitemOpen
	\bibfield  {author} {\bibinfo {author} {\bibfnamefont {A.}~\bibnamefont
			{Tero}}, \bibinfo {author} {\bibfnamefont {S.}~\bibnamefont {Takagi}},
		\bibinfo {author} {\bibfnamefont {T.}~\bibnamefont {Saigusa}}, \bibinfo
		{author} {\bibfnamefont {K.}~\bibnamefont {Ito}}, \bibinfo {author}
		{\bibfnamefont {D.~P.}\ \bibnamefont {Bebber}}, \bibinfo {author}
		{\bibfnamefont {M.~D.}\ \bibnamefont {Fricker}}, \bibinfo {author}
		{\bibfnamefont {K.}~\bibnamefont {Yumiki}}, \bibinfo {author} {\bibfnamefont
			{R.}~\bibnamefont {Kobayashi}},\ and\ \bibinfo {author} {\bibfnamefont
			{T.}~\bibnamefont {Nakagaki}},\ }\bibfield  {title} {\bibinfo {title} {Rules
			for biologically inspired adaptive network design},\ }\href@noop {}
	{\bibfield  {journal} {\bibinfo  {journal} {Science}\ }\textbf {\bibinfo
			{volume} {327}},\ \bibinfo {pages} {439} (\bibinfo {year}
		{2010})}\BibitemShut {NoStop}%
	\bibitem [{\citenamefont {Ruiz-Garc\'{\i}a}\ and\ \citenamefont
		{Katifori}(2021)}]{PhysRevE.103.062301}%
	\BibitemOpen
	\bibfield  {author} {\bibinfo {author} {\bibfnamefont {M.}~\bibnamefont
			{Ruiz-Garc\'{\i}a}}\ and\ \bibinfo {author} {\bibfnamefont {E.}~\bibnamefont
			{Katifori}},\ }\bibfield  {title} {\bibinfo {title} {Emergent dynamics in
			excitable flow systems},\ }\href@noop {} {\bibfield  {journal} {\bibinfo
			{journal} {Phys. Rev. E}\ }\textbf {\bibinfo {volume} {103}},\ \bibinfo
		{pages} {062301} (\bibinfo {year} {2021})}\BibitemShut {NoStop}%
	\bibitem [{\citenamefont {Schaub}\ and\ \citenamefont
		{Segarra}(2018)}]{Schaub:816290}%
	\BibitemOpen
	\bibfield  {author} {\bibinfo {author} {\bibfnamefont {M.~T.}\ \bibnamefont
			{Schaub}}\ and\ \bibinfo {author} {\bibfnamefont {S.}~\bibnamefont
			{Segarra}},\ }\bibfield  {title} {\bibinfo {title} {Flow smoothing and
			denoising: Graph signal processing in the edge-space},\ }\href@noop {}
	{\bibfield  {journal} {\bibinfo  {journal} {IEEE}\ ,\ \bibinfo {pages} {735}}
		(\bibinfo {year} {2018})}\BibitemShut {NoStop}%
	\bibitem [{\citenamefont {Carletti}\ \emph {et~al.}(2020)\citenamefont
		{Carletti}, \citenamefont {Battiston}, \citenamefont {Cencetti},\ and\
		\citenamefont {Fanelli}}]{PhysRevE.101.022308}%
	\BibitemOpen
	\bibfield  {author} {\bibinfo {author} {\bibfnamefont {T.}~\bibnamefont
			{Carletti}}, \bibinfo {author} {\bibfnamefont {F.}~\bibnamefont {Battiston}},
		\bibinfo {author} {\bibfnamefont {G.}~\bibnamefont {Cencetti}},\ and\
		\bibinfo {author} {\bibfnamefont {D.}~\bibnamefont {Fanelli}},\ }\bibfield
	{title} {\bibinfo {title} {Random walks on hypergraphs},\ }\href@noop {}
	{\bibfield  {journal} {\bibinfo  {journal} {Phys. Rev. E}\ }\textbf {\bibinfo
			{volume} {101}},\ \bibinfo {pages} {022308} (\bibinfo {year}
		{2020})}\BibitemShut {NoStop}%
	\bibitem [{\citenamefont {Neuh\"auser}\ \emph {et~al.}(2020)\citenamefont
		{Neuh\"auser}, \citenamefont {Mellor},\ and\ \citenamefont
		{Lambiotte}}]{PhysRevE.101.032310}%
	\BibitemOpen
	\bibfield  {author} {\bibinfo {author} {\bibfnamefont {L.}~\bibnamefont
			{Neuh\"auser}}, \bibinfo {author} {\bibfnamefont {A.}~\bibnamefont
			{Mellor}},\ and\ \bibinfo {author} {\bibfnamefont {R.}~\bibnamefont
			{Lambiotte}},\ }\bibfield  {title} {\bibinfo {title} {Multibody interactions
			and nonlinear consensus dynamics on networked systems},\ }\href@noop {}
	{\bibfield  {journal} {\bibinfo  {journal} {Phys. Rev. E}\ }\textbf {\bibinfo
			{volume} {101}},\ \bibinfo {pages} {032310} (\bibinfo {year}
		{2020})}\BibitemShut {NoStop}%
	\bibitem [{\citenamefont {Schaub}\ \emph {et~al.}(2020)\citenamefont {Schaub},
		\citenamefont {Benson}, \citenamefont {Horn}, \citenamefont {Lippner},\ and\
		\citenamefont {Jadbabaie}}]{doi:10.1137/18M1201019}%
	\BibitemOpen
	\bibfield  {author} {\bibinfo {author} {\bibfnamefont {M.~T.}\ \bibnamefont
			{Schaub}}, \bibinfo {author} {\bibfnamefont {A.~R.}\ \bibnamefont {Benson}},
		\bibinfo {author} {\bibfnamefont {P.}~\bibnamefont {Horn}}, \bibinfo {author}
		{\bibfnamefont {G.}~\bibnamefont {Lippner}},\ and\ \bibinfo {author}
		{\bibfnamefont {A.}~\bibnamefont {Jadbabaie}},\ }\bibfield  {title} {\bibinfo
		{title} {Random walks on simplicial complexes and the normalized hodge
			1-laplacian},\ }\href@noop {} {\bibfield  {journal} {\bibinfo  {journal}
			{SIAM Review}\ }\textbf {\bibinfo {volume} {62}},\ \bibinfo {pages} {353}
		(\bibinfo {year} {2020})}\BibitemShut {NoStop}%
	\bibitem [{\citenamefont {Bianconi}\ \emph {et~al.}(2019)\citenamefont
		{Bianconi}, \citenamefont {Kryven},\ and\ \citenamefont
		{Ziff}}]{PhysRevE.100.062311}%
	\BibitemOpen
	\bibfield  {author} {\bibinfo {author} {\bibfnamefont {G.}~\bibnamefont
			{Bianconi}}, \bibinfo {author} {\bibfnamefont {I.}~\bibnamefont {Kryven}},\
		and\ \bibinfo {author} {\bibfnamefont {R.~M.}\ \bibnamefont {Ziff}},\
	}\bibfield  {title} {\bibinfo {title} {Percolation on branching simplicial
			and cell complexes and its relation to interdependent percolation},\
	}\href@noop {} {\bibfield  {journal} {\bibinfo  {journal} {Phys. Rev. E}\
		}\textbf {\bibinfo {volume} {100}},\ \bibinfo {pages} {062311} (\bibinfo
		{year} {2019})}\BibitemShut {NoStop}%
	\bibitem [{\citenamefont {Alvarez-Rodriguez}\ \emph {et~al.}(2021)\citenamefont
		{Alvarez-Rodriguez}, \citenamefont {Battiston}, \citenamefont {de~Arruda},
		\citenamefont {Moreno}, \citenamefont {Perc},\ and\ \citenamefont
		{Latora}}]{Alvarez-Rodriguez2021}%
	\BibitemOpen
	\bibfield  {author} {\bibinfo {author} {\bibfnamefont {U.}~\bibnamefont
			{Alvarez-Rodriguez}}, \bibinfo {author} {\bibfnamefont {F.}~\bibnamefont
			{Battiston}}, \bibinfo {author} {\bibfnamefont {G.~F.}\ \bibnamefont
			{de~Arruda}}, \bibinfo {author} {\bibfnamefont {Y.}~\bibnamefont {Moreno}},
		\bibinfo {author} {\bibfnamefont {M.}~\bibnamefont {Perc}},\ and\ \bibinfo
		{author} {\bibfnamefont {V.}~\bibnamefont {Latora}},\ }\bibfield  {title}
	{\bibinfo {title} {Evolutionary dynamics of higher-order interactions in
			social networks},\ }\href@noop {} {\bibfield  {journal} {\bibinfo  {journal}
			{Nature Human Behaviour}\ }\textbf {\bibinfo {volume} {5}},\ \bibinfo {pages}
		{586} (\bibinfo {year} {2021})}\BibitemShut {NoStop}%
	\bibitem [{\citenamefont {Mill\'an}\ \emph {et~al.}(2020)\citenamefont
		{Mill\'an}, \citenamefont {Torres},\ and\ \citenamefont
		{Bianconi}}]{PhysRevLett.124.218301}%
	\BibitemOpen
	\bibfield  {author} {\bibinfo {author} {\bibfnamefont {A.~P.}\ \bibnamefont
			{Mill\'an}}, \bibinfo {author} {\bibfnamefont {J.~J.}\ \bibnamefont
			{Torres}},\ and\ \bibinfo {author} {\bibfnamefont {G.}~\bibnamefont
			{Bianconi}},\ }\bibfield  {title} {\bibinfo {title} {Explosive higher-order
			kuramoto dynamics on simplicial complexes},\ }\href@noop {} {\bibfield
		{journal} {\bibinfo  {journal} {Phys. Rev. Lett.}\ }\textbf {\bibinfo
			{volume} {124}},\ \bibinfo {pages} {218301} (\bibinfo {year}
		{2020})}\BibitemShut {NoStop}%
	\bibitem [{\citenamefont {Courtney}\ and\ \citenamefont
		{Bianconi}(2016)}]{PhysRevE.93.062311}%
	\BibitemOpen
	\bibfield  {author} {\bibinfo {author} {\bibfnamefont {O.~T.}\ \bibnamefont
			{Courtney}}\ and\ \bibinfo {author} {\bibfnamefont {G.}~\bibnamefont
			{Bianconi}},\ }\bibfield  {title} {\bibinfo {title} {Generalized network
			structures: The configuration model and the canonical ensemble of simplicial
			complexes},\ }\href@noop {} {\bibfield  {journal} {\bibinfo  {journal} {Phys.
				Rev. E}\ }\textbf {\bibinfo {volume} {93}},\ \bibinfo {pages} {062311}
		(\bibinfo {year} {2016})}\BibitemShut {NoStop}%
	\bibitem [{\citenamefont {Tyloo}\ \emph {et~al.}(2018)\citenamefont {Tyloo},
		\citenamefont {Coletta},\ and\ \citenamefont
		{Jacquod}}]{PhysRevLett.120.084101}%
	\BibitemOpen
	\bibfield  {author} {\bibinfo {author} {\bibfnamefont {M.}~\bibnamefont
			{Tyloo}}, \bibinfo {author} {\bibfnamefont {T.}~\bibnamefont {Coletta}},\
		and\ \bibinfo {author} {\bibfnamefont {P.}~\bibnamefont {Jacquod}},\
	}\bibfield  {title} {\bibinfo {title} {Robustness of synchrony in complex
			networks and generalized kirchhoff indices},\ }\href@noop {} {\bibfield
		{journal} {\bibinfo  {journal} {Phys. Rev. Lett.}\ }\textbf {\bibinfo
			{volume} {120}},\ \bibinfo {pages} {084101} (\bibinfo {year}
		{2018})}\BibitemShut {NoStop}%
	\bibitem [{\citenamefont {Grady}\ and\ \citenamefont
		{Polimeni}(2010)}]{grady2010discrete}%
	\BibitemOpen
	\bibfield  {author} {\bibinfo {author} {\bibfnamefont {L.}~\bibnamefont
			{Grady}}\ and\ \bibinfo {author} {\bibfnamefont {J.}~\bibnamefont
			{Polimeni}},\ }\href@noop {} {\emph {\bibinfo {title} {Discrete Calculus:
				Applied Analysis on Graphs for Computational Science}}}\ (\bibinfo
	{publisher} {Springer London},\ \bibinfo {year} {2010})\BibitemShut {NoStop}%
	\bibitem [{\citenamefont {Hirani}(2003)}]{Hirani2003DiscreteEC}%
	\BibitemOpen
	\bibfield  {author} {\bibinfo {author} {\bibfnamefont {A.~N.}\ \bibnamefont
			{Hirani}},\ }\bibfield  {title} {\bibinfo {title} {Discrete exterior
			calculus},\ }\href@noop {} {\bibfield  {journal} {\bibinfo  {journal} {Ph.D.
				dissertation, California Institute Technol., Pasadena, CA, USA}\ } (\bibinfo
		{year} {2003})}\BibitemShut {NoStop}%
	\bibitem [{\citenamefont {Lim}(2020)}]{doi:10.1137/18M1223101}%
	\BibitemOpen
	\bibfield  {author} {\bibinfo {author} {\bibfnamefont {L.-H.}\ \bibnamefont
			{Lim}},\ }\bibfield  {title} {\bibinfo {title} {Hodge laplacians on graphs},\
	}\href@noop {} {\bibfield  {journal} {\bibinfo  {journal} {SIAM Review}\
		}\textbf {\bibinfo {volume} {62}},\ \bibinfo {pages} {685} (\bibinfo {year}
		{2020})}\BibitemShut {NoStop}%
	\bibitem [{\citenamefont {Ghrist}(2014)}]{ghrist2014elementary}%
	\BibitemOpen
	\bibfield  {author} {\bibinfo {author} {\bibfnamefont {R.}~\bibnamefont
			{Ghrist}},\ }\href@noop {} {\emph {\bibinfo {title} {Elementary Applied
				Topology}}}\ (\bibinfo  {publisher} {CreateSpace Independent Publishing
		Platform},\ \bibinfo {year} {2014})\BibitemShut {NoStop}%
	\bibitem [{\citenamefont {Eckmann}(1944)}]{Eckmann1944}%
	\BibitemOpen
	\bibfield  {author} {\bibinfo {author} {\bibfnamefont {B.}~\bibnamefont
			{Eckmann}},\ }\bibfield  {title} {\bibinfo {title} {Harmonische funktionen
			und randwertaufgaben in einem komplex},\ }\href@noop {} {\bibfield  {journal}
		{\bibinfo  {journal} {Commentarii Mathematici Helvetici}\ }\textbf {\bibinfo
			{volume} {17}},\ \bibinfo {pages} {240} (\bibinfo {year} {1944})}\BibitemShut
	{NoStop}%
	\bibitem [{\citenamefont {Ben-Israel}\ and\ \citenamefont
		{Greville}(2006)}]{ben2006generalized}%
	\BibitemOpen
	\bibfield  {author} {\bibinfo {author} {\bibfnamefont {A.}~\bibnamefont
			{Ben-Israel}}\ and\ \bibinfo {author} {\bibfnamefont {T.}~\bibnamefont
			{Greville}},\ }\href@noop {} {\emph {\bibinfo {title} {Generalized Inverses:
				Theory and Applications}}},\ CMS Books in Mathematics\ (\bibinfo  {publisher}
	{Springer New York},\ \bibinfo {year} {2006})\BibitemShut {NoStop}%
	\bibitem [{\citenamefont {Arnaudon}\ \emph {et~al.}(2022)\citenamefont
		{Arnaudon}, \citenamefont {Peach}, \citenamefont {Petri},\ and\ \citenamefont
		{Expert}}]{arnaudon2022connecting}%
	\BibitemOpen
	\bibfield  {author} {\bibinfo {author} {\bibfnamefont {A.}~\bibnamefont
			{Arnaudon}}, \bibinfo {author} {\bibfnamefont {R.~L.}\ \bibnamefont {Peach}},
		\bibinfo {author} {\bibfnamefont {G.}~\bibnamefont {Petri}},\ and\ \bibinfo
		{author} {\bibfnamefont {P.}~\bibnamefont {Expert}},\ }\bibfield  {title}
	{\bibinfo {title} {Connecting hodge and sakaguchi-kuramoto through a
			mathematical framework for coupled oscillators on simplicial complexes},\
	}\href@noop {} {\bibfield  {journal} {\bibinfo  {journal} {Communications
				Physics}\ }\textbf {\bibinfo {volume} {5}},\ \bibinfo {pages} {211} (\bibinfo
		{year} {2022})}\BibitemShut {NoStop}%
	\bibitem [{\citenamefont {Nurisso}\ \emph {et~al.}(2024)\citenamefont
		{Nurisso}, \citenamefont {Arnaudon}, \citenamefont {Lucas}, \citenamefont
		{Peach}, \citenamefont {Expert}, \citenamefont {Vaccarino},\ and\
		\citenamefont {Petri}}]{nurisso2024unified}%
	\BibitemOpen
	\bibfield  {author} {\bibinfo {author} {\bibfnamefont {M.}~\bibnamefont
			{Nurisso}}, \bibinfo {author} {\bibfnamefont {A.}~\bibnamefont {Arnaudon}},
		\bibinfo {author} {\bibfnamefont {M.}~\bibnamefont {Lucas}}, \bibinfo
		{author} {\bibfnamefont {R.~L.}\ \bibnamefont {Peach}}, \bibinfo {author}
		{\bibfnamefont {P.}~\bibnamefont {Expert}}, \bibinfo {author} {\bibfnamefont
			{F.}~\bibnamefont {Vaccarino}},\ and\ \bibinfo {author} {\bibfnamefont
			{G.}~\bibnamefont {Petri}},\ }\bibfield  {title} {\bibinfo {title} {A unified
			framework for simplicial kuramoto models},\ }\href@noop {} {\bibfield
		{journal} {\bibinfo  {journal} {Chaos: An Interdisciplinary Journal of
				Nonlinear Science}\ }\textbf {\bibinfo {volume} {34}},\ \bibinfo {pages}
		{053118} (\bibinfo {year} {2024})}\BibitemShut {NoStop}%
	\bibitem [{\citenamefont {Strogatz}\ and\ \citenamefont
		{Mirollo}(1991)}]{Strogatz1991}%
	\BibitemOpen
	\bibfield  {author} {\bibinfo {author} {\bibfnamefont {S.~H.}\ \bibnamefont
			{Strogatz}}\ and\ \bibinfo {author} {\bibfnamefont {R.~E.}\ \bibnamefont
			{Mirollo}},\ }\bibfield  {title} {\bibinfo {title} {Stability of incoherence
			in a population of coupled oscillators},\ }\href@noop {} {\bibfield
		{journal} {\bibinfo  {journal} {Journal of Statistical Physics}\ }\textbf
		{\bibinfo {volume} {63}},\ \bibinfo {pages} {613} (\bibinfo {year}
		{1991})}\BibitemShut {NoStop}%
	\bibitem [{\citenamefont {Ghorbanchian}\ \emph {et~al.}(2021)\citenamefont
		{Ghorbanchian}, \citenamefont {Restrepo}, \citenamefont {Torres},\ and\
		\citenamefont {Bianconi}}]{Ghorbanchian2021}%
	\BibitemOpen
	\bibfield  {author} {\bibinfo {author} {\bibfnamefont {R.}~\bibnamefont
			{Ghorbanchian}}, \bibinfo {author} {\bibfnamefont {J.~G.}\ \bibnamefont
			{Restrepo}}, \bibinfo {author} {\bibfnamefont {J.~J.}\ \bibnamefont
			{Torres}},\ and\ \bibinfo {author} {\bibfnamefont {G.}~\bibnamefont
			{Bianconi}},\ }\bibfield  {title} {\bibinfo {title} {Higher-order simplicial
			synchronization of coupled topological signals},\ }\href@noop {} {\bibfield
		{journal} {\bibinfo  {journal} {Communications Physics}\ }\textbf {\bibinfo
			{volume} {4}},\ \bibinfo {pages} {120} (\bibinfo {year} {2021})}\BibitemShut
	{NoStop}%
	\bibitem [{\citenamefont {Kuramoto}(1984)}]{Kuramoto1984}%
	\BibitemOpen
	\bibfield  {author} {\bibinfo {author} {\bibfnamefont {Y.}~\bibnamefont
			{Kuramoto}},\ }\bibinfo {title} {Mutual entrainment},\ in\ \href@noop {}
	{\emph {\bibinfo {booktitle} {Chemical Oscillations, Waves, and
				Turbulence}}}\ (\bibinfo  {publisher} {Springer Berlin Heidelberg},\ \bibinfo
	{address} {Berlin, Heidelberg},\ \bibinfo {year} {1984})\ pp.\ \bibinfo
	{pages} {60--88}\BibitemShut {NoStop}%
	\bibitem [{\citenamefont {Zhu}\ \emph {et~al.}(1996)\citenamefont {Zhu},
		\citenamefont {Klein},\ and\ \citenamefont {Lukovits}}]{Zhu1996}%
	\BibitemOpen
	\bibfield  {author} {\bibinfo {author} {\bibfnamefont {H.-Y.}\ \bibnamefont
			{Zhu}}, \bibinfo {author} {\bibfnamefont {D.~J.}\ \bibnamefont {Klein}},\
		and\ \bibinfo {author} {\bibfnamefont {I.}~\bibnamefont {Lukovits}},\
	}\bibfield  {title} {\bibinfo {title} {Extensions of the wiener number},\
	}\href@noop {} {\bibfield  {journal} {\bibinfo  {journal} {Journal of
				Chemical Information and Computer Sciences}\ }\textbf {\bibinfo {volume}
			{36}},\ \bibinfo {pages} {420} (\bibinfo {year} {1996})}\BibitemShut
	{NoStop}%
	\bibitem [{\citenamefont {Gutman}\ and\ \citenamefont
		{Mohar}(1996)}]{Gutman1996}%
	\BibitemOpen
	\bibfield  {author} {\bibinfo {author} {\bibfnamefont {I.}~\bibnamefont
			{Gutman}}\ and\ \bibinfo {author} {\bibfnamefont {B.}~\bibnamefont {Mohar}},\
	}\bibfield  {title} {\bibinfo {title} {The quasi-wiener and the kirchhoff
			indices coincide},\ }\href@noop {} {\bibfield  {journal} {\bibinfo  {journal}
			{Journal of Chemical Information and Computer Sciences}\ }\textbf {\bibinfo
			{volume} {36}},\ \bibinfo {pages} {982} (\bibinfo {year} {1996})}\BibitemShut
	{NoStop}%
\end{thebibliography}

%

\end{document}